\documentclass[prd,twocolumn,showpacs,amsmath,amssymb]{revtex4}
\usepackage{graphicx}
\usepackage{dcolumn}
\usepackage{bm}
\begin{document}
\title{A Reinvestigation of Moving Punctured Black Holes with a New Code}

\author{Zhoujian Cao\footnote{zjcao@amt.ac.cn}}

\affiliation{Institute of Applied Mathematics, Academy of Mathematics and
System Science, Chinese Academy of Sciences, Beijing 100080, China,\\
Theoretical Institute for Advanced Research in Astrophysics,
Academia Sinica, Taiwan,\\
State Key Laboratory of Scientific and Engineering Computing, China}
\author{Hwei-Jang Yo\footnote{hjyo@phys.ncku.edu.tw} and Jui-Ping Yu}
\affiliation{Department of
Physics, National Cheng-Kung University, Tainan 701, Taiwan}
\begin{abstract}
We report on our code, in which the moving puncture method is
applied and an adaptive/fixed mesh refinement is implemented, and
on its preliminary performance on black hole simulations. Based on
the BSSN formulation, up-to-date gauge conditions and the
modifications of the formulation are also implemented and tested.
In this work we present our primary results about the simulation
of a single static black hole, of a moving single black hole, and
of the head-on collision of a binary black hole system. For the
static punctured black hole simulations, different modifications
of the BSSN formulation are applied. It is demonstrated that both
the currently used sets of modifications lead to a stable
evolution. For cases of a moving punctured black hole with or
without spin, we search for viable gauge conditions and study the
effect of spin on the black hole evolution. Our results confirm
previous results obtained by other research groups. In addition,
we find a new gauge condition, which has not yet been adopted by
any other researchers, which can also give stable and accurate
black hole evolution calculations. We examine the performance of
the code for the head-on collision of a binary black hole system,
and the agreement of the gravitational waveform it produces with
that obtained in other works.
In order to understand qualitatively the influence of matter on 
the binary black hole collisions, we also investigate the same
head-on collision scenarios but perturbed by a scalar field.
The numerical simulations performed
with this code not only give stable and accurate results that are
consistent with the works by other numerical relativity groups,
but also lead to the discovery of a new viable gauge condition,
as well as clarify some ambiguities in the modification of the BSSN
formulation. These results demonstrate that this code is reliable
and ready to be used in the study of more realistic astrophysical
scenarios and of numerical relativity.
\end{abstract}
\date{\today}

\maketitle
%%%%%%%%%%%%%%%%%%%%%%%%%%%%%%%%%%%%%%%%%%%%%%%%%%%%%%%%%%%%%%%%%%
\section{introduction}
%%%%%%%%%%%%%%%%%%%%%%%%%%%%%%%%%%%%%%%%%%%%%%%%%%%%%%%%%%%%%%%%%%
There are two main purposes for studying numerical relativity: One
is to investigate the mathematical issues of the geometry of the
Einstein manifold. These issues include cosmic censorship, the
hoop conjecture, the Penrose inequality, and critical phenomenon
\cite{berger02}. The other, more practical one, is to study the
dynamics of astrophysically compact objects. To meet the needs of
existing (e.g., LIGO \cite{LIGO}, VIRGO \cite{VIRGO}, GEO600
\cite{GEO}, and TAMA \cite{TAMA}) and planned (e.g. LISA
\cite{LISA}) gravitational wave detectors, the theoretical
prediction of the gravitational waveform for realistic sources
has become urgent. In both of the aspects of numerical relativity,
the most important targets for study are black holes and neutron
stars, especially those in binary systems.

After decades of exploration by many researchers, recently there
have been exciting breakthroughs in the simulation of the 
evolution of binary black holes (BBHs)
\cite{bbhsuccess,moving_punctur1,moving_punctur2}. Soon after
these breakthroughs, the moving puncture method based on the BSSN
formalism \cite{modified_3+1} was widely used by numerical
relativity groups to deal with black hole systems
\cite{vaishnav07,SpeU07,brugmann08} and neutron star systems
\cite{binary_netron_star}. Much interesting physics related to
BBHs has been explored, including the waveform of the
gravitational radiation \cite{waveform,baker07c}, the spin-orbit
coupling effect \cite{spin_orbit_coupling}, and the recoil
velocity \cite{recoil,campanelli07c}. It has been emphasized in
almost all of these studies that the treatment of the singularity
problem, i.e., the moving puncture method and the gauge condition,
are both key factors in this series of successes. Not needing any
inner boundary condition makes the moving puncture method much
simpler to handle, compared with the excision method
\cite{mpbhana}. This simplicity has made the moving puncture
method popular in the numerical relativity community. The
Bona-Masso type slicing gauge conditions \cite{bona95} for the
lapse function and many driver gauge conditions (e.g., the
$\Gamma$-driver) for the shift vector \cite{balakrishna96,meter06}
have been shown to be very important in making the moving puncture
method work. However, in \cite{meter06} it was shown that,
although the details of the gauge conditions used in the punctured
BBH evolutions are different, only certain gauge choices allow one
to evolve a single moving puncture black hole. It is desirable
to better understand the effect of the gauge choices
on black hole evolutions.

In order to study numerical relativity and also
investigate the specific aforementioned topics we develop,
from scratch, a code based on the moving puncture method. We adopt a
fourth-order finite differencing scheme for the spatial
derivatives and the Crank-Nicholson scheme for the time
integrations. We use the GrACE package \cite{GrACE} to implement
both the mesh refinement and the parallelization in our code. In
this paper we present our primary results about the evolution of a
single static black hole, of a single moving black hole, of the
head-on collision of a BBH, and of the head-on collision perturbed
by a scalar field. Our results confirm many results obtained in
the previous works by other groups. On the other hand we find a
new gauge condition, which has not been tried by other
researchers, that can also give stable and accurate black hole
evolution calculations. We also observe the effects of the
existence of a massless scalar field in delaying the head-on
collision, depending on the initial configuration of the scalar
field. All of these results enhance our confidence in this code,
and thus we will apply the code to more realistic astrophysical
calculations in the near future.

The remainder of the paper is organized as follows: In Section
\ref{secii}, we first summarize the BSSN formulation, the
conventional modifications and adjustments, and the gauge
conditions. Then we describe the numerical methods used in this
code in section \ref{seciii}, including the details of the FMR/AMR
algorithm, the finite-differencing stencils, and the Kreiss-Oliger
dissipation. In Section \ref{seciv}, the initial data is outlined.
In Section \ref{secv}, we present our numerical results on the
evolutions of a single static black hole, of a single moving black
hole with and without spin, and of the head-on collision of a BBH,
with and without a massless scalar field. We summarize and discuss
the implications of our findings in Sec. \ref{secvi}.
%%%%%%%%%%%%%%%%%%%%%%%%%%%%%%%%%%%%%%%%%%%%%%%%%%%%%%%%%%%%%%%%%%%
\section{formulation} \label{secii}
%%%%%%%%%%%%%%%%%%%%%%%%%%%%%%%%%%%%%%%%%%%%%%%%%%%%%%%%%%%%%%%%%%%
\subsection{Review of the Basic Equations}
%%%%%%%%%%%%%%%%%%%%%%%%%%%%%%%%%%%%%%%%%%%%%%%%%%%%%%%%%%%%%%%%%%
The code is based on the BSSN formalism \cite{modified_3+1}, which
is a conformal-traceless ``3+1" formulation of the Einstein
equations. In this formalism, the spacetime is decomposed into
three-dimensional spacelike slices, described by a three-metric
$\gamma_{ij}$; its embedding in the four-dimensional spacetime is
specified by the extrinsic curvature $K_{ij}$ and the variables,
the lapse $\alpha$ and shift vector $\beta^i$, that specify a
coordinate system.  Our conventions are that Latin indices run
over 1, 2, 3, whereas Greek indices run over 0, 1, 2, 3.
Throughout the paper we adopt geometrical units with $G=c=1$. In
this paper we follow the notations of \cite{baumgarte03}. The
metric $\gamma_{ij}$ is conformally transformed via
\begin{equation}
\phi\equiv\frac{1}{12} \ln \gamma, \,\,\, \tilde{\gamma}_{ij}\equiv
e^{-4\phi} \gamma_{ij},
\end{equation}
where $\gamma$ denotes the determinant of the metric
$\gamma_{ij}$. The conformal exponent $\phi$ is evolved as an
independent variable, whereas $\tilde{\gamma}_{ij}$ is subjected
to the constraint that the determinant of $\tilde{\gamma}_{ij}$ is
chosen to be unimodular, i.e., $\tilde{\gamma}=1$. The extrinsic
curvature is subjected to the same conformal transformation, and
its trace $K$ is evolved as an independent variable. That is, in
place of $K_{ij}$ we evolve:
\begin{equation}
K\equiv\gamma^{ij}K_{ij}, \,\,\, \tilde{A}_{ij}\equiv e^{-4\phi}
K_{ij}-\frac{1}{3}\tilde\gamma_{ij} K. \label{defineA}
\end{equation}
Similar to the conformal metric, $\tilde{A}_{ij}$ is subjected to
a constraint that $\tilde{A}_{ij}$ is traceless, i.e.,
tr$\tilde{A}_{ij}=0$. New evolution variables, i.e., the conformal
connections,
\begin{equation}
\tilde{\Gamma}^i\equiv-\tilde{\gamma}^{ij}{}_{,j},
\end{equation}
are introduced, defined in terms of the contraction of the spatial derivative
of the inverse conformal three-metric ${\tilde\gamma}^{ij}$.

With these dynamical variables the evolution equations read
\begin{eqnarray}
\partial_t \phi &=& \beta^i \phi_{,i} - \frac{1}{6} \alpha K
     + \frac{1}{6} \beta^i{}_{,i},\label{eq4}\\
\partial_t\tilde\gamma_{ij} &=&\beta^k\tilde{\gamma}_{ij,k}
    - 2 \alpha \tilde A_{ij}+2\tilde\gamma_{k(i} \beta^k{}_{,j)}
    - \frac{2}{3} \tilde \gamma_{ij}\beta^k{}_{,k},\label{dtgij}\\
\partial_t K&=&\beta^i K_{,i}-D^2\alpha +
    \alpha[\tilde A_{ij} \tilde A^{ij}+\frac{1}{3} K^2+4\pi(\rho + s)],
    \nonumber\\ &&\label{dtK}\\
\partial_t\tilde A_{ij}& = &\beta^k \tilde A_{ij,k} + e^{- 4 \phi}
 [\alpha(R_{ij}-8\pi s_{ij} )-D_iD_j\alpha]^{TF}  \nonumber\\
    &+&\alpha (K \tilde A_{ij} - 2\tilde A_{ik}\tilde A^k_j)
 +2\tilde A_{k(i} \beta^k{}_{,j)}-\frac{2}{3}\tilde A_{ij}\beta^k{}_{,k},
\nonumber\\
 &&\label{dtAij}\\
\partial_t \tilde{\Gamma}^i &=& \beta^j \tilde \Gamma^i{}_{,j}
    - 2 \tilde A^{ij} \alpha_{,j} \nonumber\\
&+& 2 \alpha \Big( \tilde \Gamma^i_{jk} \tilde A^{kj}
    - \frac{2}{3} \tilde \gamma^{ij} K_{,j}
   - 8 \pi \tilde \gamma^{ij} s_j
    + 6 \tilde A^{ij} \phi_{,j} \Big) \nonumber\\
    & -& \tilde \Gamma^j \beta^i{}_{,j}
    + \frac{2}{3} \tilde \Gamma^i \beta^j{}_{,j}
    + \frac{1}{3} \tilde \gamma^{ki} \beta^j_{,jk}
    + \tilde \gamma^{kj} \beta^i_{,kj}.\label{dtGamma}
\end{eqnarray}
Here $\rho$, $s$, $s_i$, $s_{ij}$ are source terms which come from
matter. For a vacuum spacetime $\rho=s=s_i=s_{ij}=0$. In the above
evolution equations $D_i$ is the covariant derivative associated
with the three-metric $\gamma_{ij}$, and ``TF" indicates the
trace-free part of tensor objects. The Ricci tensor $R_{ij}$ is
given as
\begin{eqnarray}
R_{ij}&=&\tilde{R}_{ij}+R_{ij}^\phi,\\
\tilde{R}_{ij}&=&-\frac{1}{2}\tilde\gamma^{mn}\tilde\gamma_{ij,mn}
+\tilde\gamma_{k(i}\tilde\Gamma^k{}_{,j)}
+\tilde\Gamma^k\tilde\Gamma_{(ij)k}\nonumber\\
&&+\tilde\gamma^{mn}(2\tilde\Gamma^k{}_{m(i}\tilde\Gamma_{j)kn}
+\tilde\Gamma^k{}_{in}\tilde\Gamma_{kmj}),\label{confricci}\\
R^\phi_{ij}&=&-2\tilde D_i\tilde D_j\phi-2\tilde\gamma_{ij}
\tilde D^k\tilde D_k\phi\nonumber\\
&&+4\tilde D_i\phi\tilde D_j\phi-4\tilde\gamma_{ij}
    \tilde D^k\phi\tilde D_k\phi .\label{ricciphi}
\end{eqnarray}
The Einstein equations also lead to a set of physical constraint
equations that are satisfied within each spacelike slice:
\begin{eqnarray}
e^{-4\phi}(\tilde{R}-8\tilde{D}^i\tilde{D}_i\phi-
8\tilde D^k\phi\tilde D_k\phi)\qquad\quad&&\nonumber\\
+\frac{2}{3}K^2-\tilde{A}_{ij}\tilde{A}^{ij}-16\pi\rho&=&0,\\
\tilde{D}^i\tilde{A}_{ij}+6\tilde{A}_{ij}\tilde{D}^i\phi-
\frac{2}{3}\tilde{D}_jK-8\pi s_j&=&0,
\end{eqnarray}
which are usually referred to as the Hamiltonian and the momentum
constraints. Here $\tilde R=\tilde R^i{}_i$ is the conformal Ricci
scalar on a three-dimensional time slice, and $\tilde{D}_i$ is the
covariant derivative associated with the conformal three-metric
${\tilde\gamma}_{ij}$. Besides being used to obtain the evolution
equations (\ref{dtK}) and (\ref{dtGamma}) in the BSSN formulation,
the Hamiltonian and the momentum constraints are also applied to
the volume integrals of the ADM mass and the angular momentum,
respectively \cite{YHBS02}:
\begin{eqnarray}
M&=&\frac{1}{16\pi}\oint_{\partial\Omega}({\tilde\Gamma}^i
-8{\tilde\gamma}^{ij}\partial_je^\phi){\rm d}{\tilde\Sigma}_i\\
&=&\frac{1}{16\pi}\int_\Omega{\rm d}^3x[e^{5\phi}
        (16\pi\rho + \tilde{A}_{ij}\tilde{A}^{ij} -\frac{2}{3}K^2)\nonumber\\
 &&\qquad\qquad\qquad\qquad\qquad\quad +\tilde{\Gamma}^k{}_{,k}
 - e^\phi\tilde{R}],\label{volmass}\\
   J_i &=&\frac{1}{8\pi} \epsilon_{ij}{}^k \oint_{\partial\Omega}
           e^{6\phi}x^j \tilde{A}^\ell{}_k{\rm d}\tilde{\Sigma}_\ell\\
    &=&\frac{1}{8\pi}\epsilon_{ij}{}^k\int_\Omega{\rm d}^3x
           [e^{6\phi}(\tilde{A}^j{}_k+\frac{2}{3}x^jK_{,k}\nonumber\\
    &&\qquad\qquad\quad- \frac{1}{2} x^j\tilde{A}_{\ell m}
     \tilde{\gamma}^{\ell m}{}_{,k}+8\pi x^js_k)],
\end{eqnarray}
where ${\rm d}\tilde{\Sigma}_i=(1/2)\epsilon_{ijk}{\rm d}x^j{\rm
d}x^k$. These two global quantities are useful tools for the system
diagnostics to validate the calculations. The volume integral
(\ref{volmass}) is slightly different from the one in
\cite{YHBS02} due to the further application of the unimodular
determinant of the conformal metric (\ref{detg1}). Refer to
Appendix \ref{unimod}
 for the details.
%%%%%%%%%%%%%%%%%%%%%%%%%%%%%%%%%%%%%%%%%%
\subsection{Equation Adjustments}\label{seciib}
%%%%%%%%%%%%%%%%%%%%%%%%%%%%%%%%%%%%%%%%%%
The specific choice of evolution variables introduces five additional
constraints,
\begin{eqnarray}
\tilde{\gamma}-1&=&0,\label{detg1}\\
\text{tr}\tilde{A}_{ij}&=&0,\label{trA0}\\
\tilde{\Gamma}^i+\tilde{\gamma}^{ij}{}_{,j}&=&0.\label{Gofg}
\end{eqnarray}
Our code actively enforces the algebraic constraints (\ref{detg1})
and (\ref{trA0}) by replacing $\tilde\gamma_{ij}$ and $\tilde
A_{ij}$ with the following:
\begin{eqnarray}
\tilde\gamma_{ij}&\rightarrow&\tilde\gamma^{-1/3}\tilde\gamma_{ij},
\label{gamave} \\
\tilde A_{ij}&\rightarrow&\tilde A_{ij}-\frac{1}{3}\tilde\gamma_{ij}{\rm tr}
\tilde A_{ij}.\label{Aave}
\end{eqnarray}
To enforce Eq.~(\ref{Gofg}) all the undifferentiated
$\tilde\Gamma^i$ in the evolution equations are substituted with
$-\tilde\gamma^{ij}{}_{,j}$.

As to the variable choice for the conformal factor, the alternative
$\chi$, first proposed in \cite{moving_punctur1}, has been widely
adopted. In the $\chi$ method the conformal exponent $\phi$ [which
is $O(\ln r)$ near the puncture] is replaced with a new variable
$\chi\equiv e^{-4\phi}$ [which is $O(r^4)$ near the puncture].
$\chi$ grows linearly near the puncture during the time evolution;
such linear behavior leads to a more accurate evolution near the
puncture. In the $\chi$ method, equation (\ref{eq4}) is replaced
by
\begin{equation}
\partial_t\chi=\frac{2}{3}\chi(\alpha K-\beta^{i}{}_{,i})+\beta^{i}\chi_{,i}.
\end{equation}
Note that $\phi_{,i}=-\chi_{,i}/4\chi$ and $\phi_{,ij}
=\chi_{,i}\chi_{,j}/4\chi^2-\chi_{,ij}/4\chi$ are applied to the
evolution equations (\ref{dtK}), (\ref{dtAij}) [via
Eq.~(\ref{ricciphi})], and (\ref{dtGamma}). In these
substitutions, the divisions by $\chi$ need to be taken care of in
the numerical implementation to avoid division by zero or
unphysically negative values of $\chi$. In \cite{confexp2} a small
$\epsilon$ is set to replace $\chi$ in division if $\chi$ is less
than $\epsilon$. In \cite{confexp1} $W\equiv e^{-2\phi}$ is chosen
to be the conformal factor variable instead of $\chi$, to avoid
the effect of unphysical negative values of $\chi$ on the
evolution of the other variables \footnote{It is also pointed out in
\cite{confexp1} that $W$ can make the numerical
computation near the black hole more accurate than both $\chi$ and $\phi$.
It is shown in \cite{shibata08} that $W$ is
more convenient than $\phi$ to compute the Ricci tensor since
$R_{ij}=\tilde{R}_{ij}+R^W_{ij}$ with $R^W_{ij}=\tilde{D}_i\tilde{D}_j
W/W+\tilde{\gamma}_{ij}(\tilde{D}_k\tilde{D}^k W/W-2\tilde{D}_k W
\tilde{D}^k W/W^2)$, which is formally simpler than Eq.~(\ref{ricciphi}).}.
However, we did not encounter any such difficulty in the work for
this paper, therefore it is not necessary to apply the
aforementioned modifications, although we anticipate the appearance
of this difficulty in some complicated scenarios in future work.

%%%%%%%%%%%%%%%%%%%%%%%%%%%%%%%%%%%%%%%%%%%%%%%%%%%%%%%%%%
\subsection{Gauge Conditions}
%%%%%%%%%%%%%%%%%%%%%%%%%%%%%%%%%%%%%%%%%%%%%%%%%%%%%%%%%%
As mentioned in the introduction, the gauge conditions are
important for the numerical simulations of dynamical spacetime, and
this is especially true for the moving puncture method. The
Bona-Masso type slicing gauge conditions \cite{bona95} for the
lapse function and many driver gauge conditions (e.g., the
$\Gamma$-driver) for the shift vector \cite{balakrishna96,meter06}
are currently the main type of gauge conditions used in the
punctured black hole calculations. In this work, we will only
focus on these types of gauge conditions, which can be written as
\begin{eqnarray}
\partial_t\alpha&=&-2\alpha K+\lambda_1\beta^i\alpha_{,i},\label{lapse_eq}\\
\partial_t\beta^i&=&\frac{3}{4}f(\alpha)B^i+\lambda_2\beta^j\beta^i{}_{,j},
\label{beta_eq}\\
\partial_t B^i&=&\partial_t\tilde{\Gamma}^i-\eta B^i+\lambda_3\beta^jB^i{}_{,j}
-\lambda_4\beta^j \tilde{\Gamma}^i{}_{,j},\label{b_eq}
\end{eqnarray}
where $\eta$ and the four $\lambda$'s are the parameters to be
chosen, and $f(\alpha)$ is a function of $\alpha$. Here the
$\lambda$'s can only be set to be 0 or 1. We set $f(\alpha)=1$ in
all the cases except in Sec.~\ref{secva}, where
$f(\alpha)=\alpha$. The gauge conditions used for moving black
holes in the literature include: (1)
$\lambda_1=1,\lambda_2=\lambda_3=\lambda_4=0$ (e.g., see
\cite{campanelli07c}); and (2)
$\lambda_1=\lambda_2=\lambda_3=\lambda_4=1$ (e.g., see
\cite{baker07c,brugmann08}), with the proper $\eta$'s. In
\cite{meter06}, the authors investigated several cases of the
above gauge equations. In \cite{gundlach06a}, the authors
discussed the above case (2) analytically. In this work, we will
explore this problem more thoroughly (see the following sections
for details). In particular, we are concerned about the effect of
the advection terms on the stability and accuracy of the
evolution, and we try to find the viable gauges for moving and/or
spinning black holes. Throughout this paper we will fix $\eta=2$.
%%%%%%%%%%%%%%%%%%%%%%%%%%%%%%%%%%%%%%%%%%%%%%%%%%%%%%%%%%
\section{numerical method}\label{seciii}
%%%%%%%%%%%%%%%%%%%%%%%%%%%%%%%%%%%%%%%%%%%%%%%%%%%%%%%%%%
In this section, we briefly describe the key numerical techniques
used in this work. For the discretization, our code uses the
cell-centered method, which takes the data to be defined at the
center of the spatial grid cell. We also use a finite
centered-differencing method with fourth-order accuracy to
approximate the spatial derivatives, which closely follows
\cite{zlochower05}. In the temporal part, we use the iterative
Crank-Nicholson method for time integration, which gives a
second-order accuracy \cite{teukolsky00}. We take d$t$=(Courant
factor)$\times$d$x$, and the Courant factor is set to be 1/4.

We apply the standard centered finite differencing approximation to
all spatial derivatives except the advection terms (i.e., the
terms of the form $\beta^{j}\partial_j F$). For these advection
terms we use the following fourth-order lop-sided stencils:
\begin{eqnarray}
\partial_x F_{i,j,k}&=&\frac{1}{12{\rm d}x}(-F_{i-3,j,k}+6F_{i-2,j,k}
                       -18F_{i-1,j,k}\nonumber \\
   &&+10F_{i,j,k}+3F_{i+1,j,k})\quad\mbox{for}\quad\beta^x<0,\\
\partial_x F_{i,j,k}&=&\frac{1}{12{\rm d}x}(F_{i+3,j,k}-6F_{i+2,j,k}
                       +18F_{i+1,j,k}\nonumber \\
   &&-10F_{i,j,k}-3F_{i-1,j,k})\quad\mbox{for}\quad\beta^x>0,
\end{eqnarray}
along the $x$-direction. The stencils are similar along the $y$-
and $z$-directions. We also install in the code a Kreiss-Oliger
dissipation \cite{KO} of the form
\begin{equation}
\partial_tF\rightarrow{\rm RHS}+\epsilon(-1)^{n/2}\sum h_i^{n+1}D_{i+}^{n/2+1}
D_{i-}^{n/2+1}F,
\end{equation}
where RHS represents the corresponding evolution equation for $F$,
$h_i$ is the grid spacing in the $i$th direction, $D_{i+}$ and
$D_{i-}$ are the forward and backward differencing operators in
the $i$th-direction, $n$ is the order of the finite difference
used to evaluate the RHS, and $\epsilon$ is the dissipation
coefficient to be chosen in various cases.

In order to increase the numerical resolution without increasing
the computation cost much, the mesh refinement method is used in
the numerical simulations. We use the GrACE package \cite{GrACE}
to implement both the mesh refinement and the parallelization in
our code. This package is able to deal with the adaptive system,
considering both partitioning and load-balancing for distributed
adaptive mesh refinement applications. However, we only use fixed
mesh refinement (FMR) in this work. The computational domain is
represented by a hierarchy of nested Cartesian grids; we adopt the
cell-centered scheme, and the grid hierarchy follows the Berger
and Oliger algorithm \cite{BO}. The hierarchy consists of $L$
levels of refinement indexed by $\ell=0$, ... , $L-1$, for which
$\ell=0$ is the coarsest level and $\ell=L-1$ is the finest one. A
refinement level consists of one or more Cartesian grids with
constant grid spacing $h_\ell$ on level $\ell$. All grid blocks
have the same logical structure, and refinements are bisected in
each coordinate direction. A refinement factor of $2$ is used such
that $h_\ell=h_0/2^\ell$. The grids are properly nested such that
the coordinate extent of any grid at level $\ell$, $\ell>0$, is
completely covered by the grids at level $\ell-1$. We do not
refine the time step. So the time step for each level is set as
d$t$ = (Courant factor) $\times$ $h_{\rm min}$, where $h_{\rm
min}$ is the resolution of the finest level. Therefore, there is
no interpolation of data between different time slices.

\begin{figure}[t]
\begin{tabular}{rl}
\includegraphics[width=0.233\textwidth]{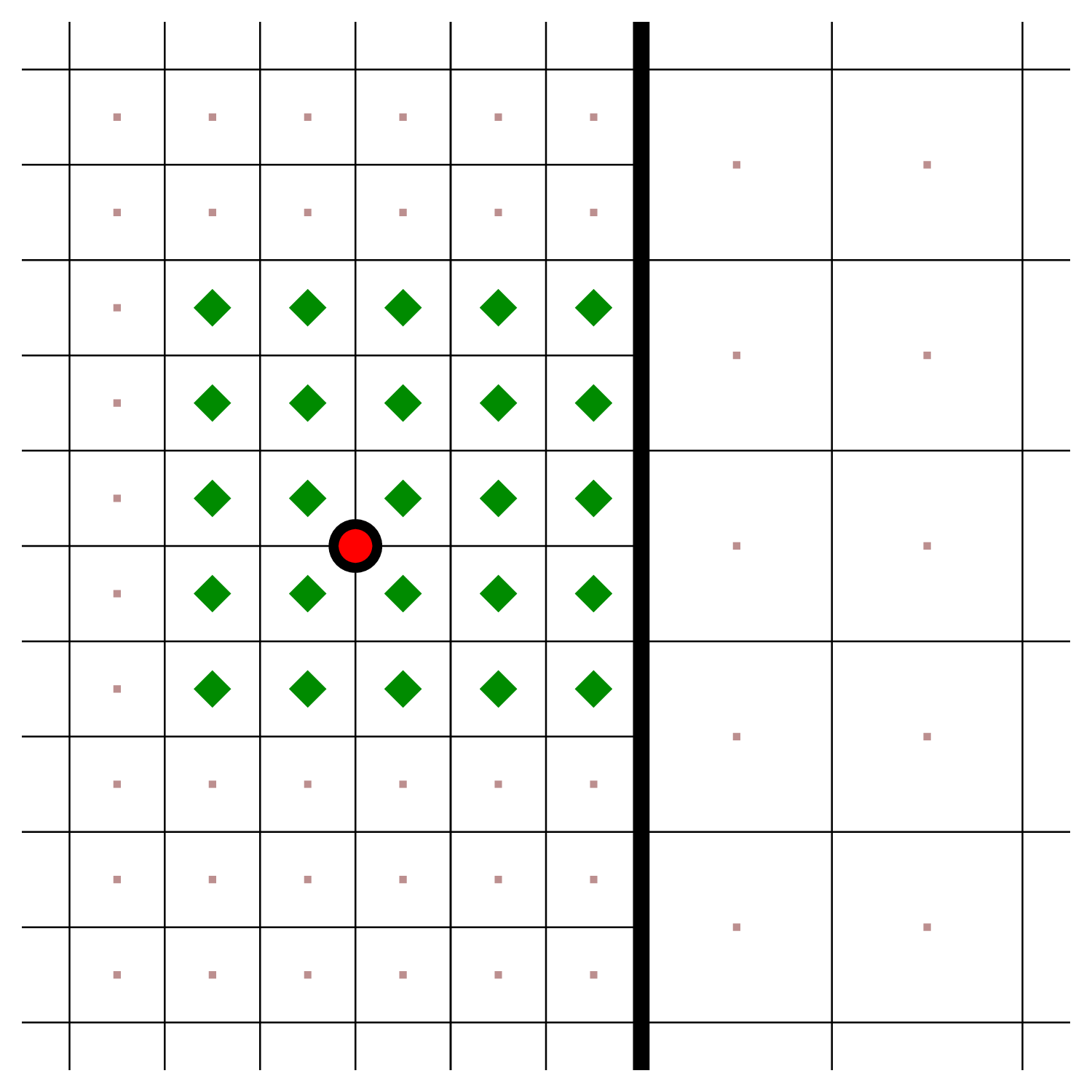}&
\includegraphics[width=0.233\textwidth]{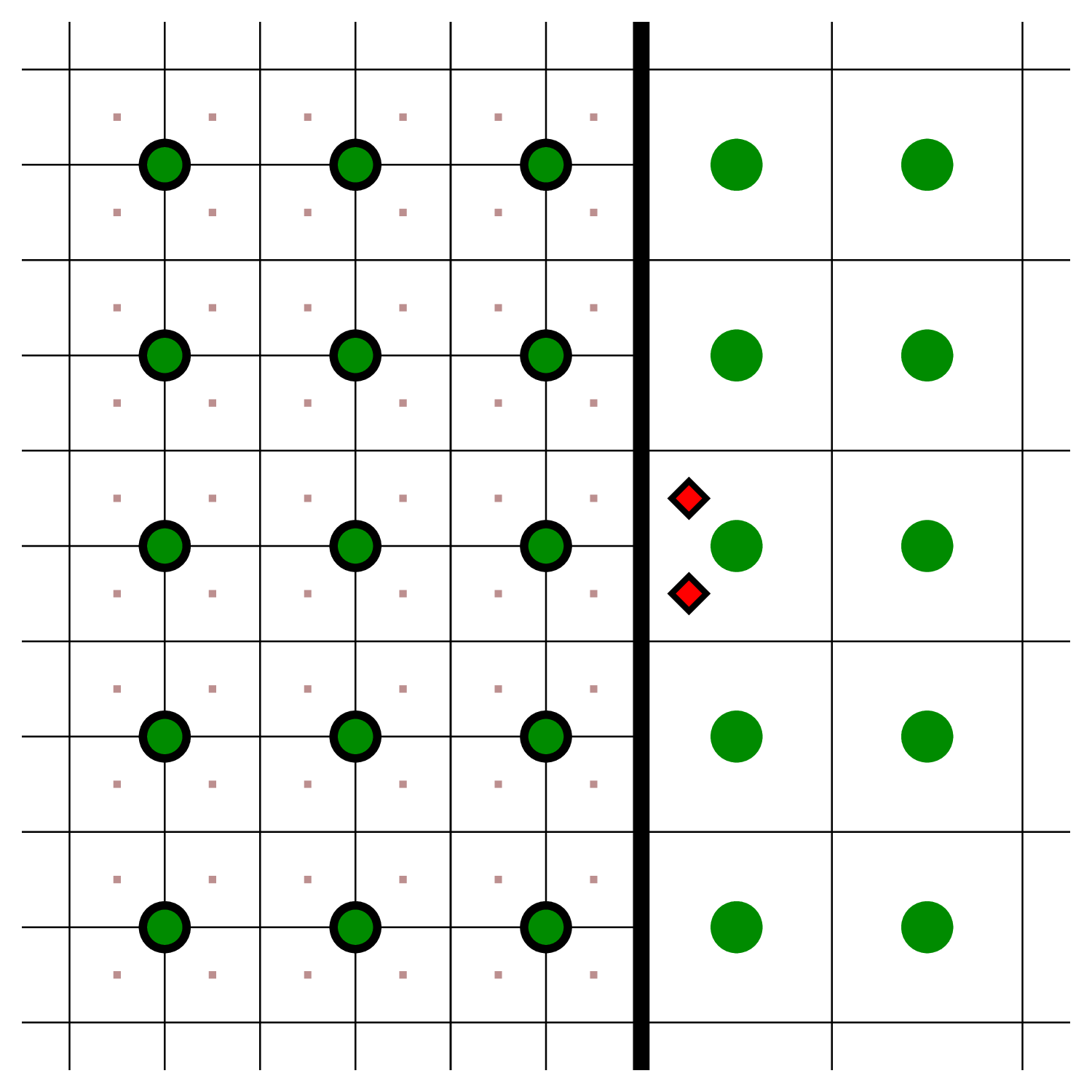}
\end{tabular}
\caption{(closely follows Fig.1 of \cite{imbiriba04})
Two-dimensional diagrams for the guard cell filling. The thick
vertical lines in both panels represent the refinement boundaries
separating fine and coarse grid regions. The left panel shows the
first step, in which one of the coarse grid cells (red circles
with black rim) are filled using a quartic interpolation across 25
interior fine grid cells (green diamonds). The right panel shows
the second step in which two fine grid guard cells (red diamonds
with black rim) are filled using quartic interpolations across 25
coarse grid values (circles). These coarse grid values include two
layers of guard cells (green circles), obtained from the coarse
grid region to the right of the interface, and three layers of
interior cells (green circles with black rims). The final step
(not shown in the figure) is to use ``derivative matching'' to
fill the guard cells for the coarse grid.} \label{guard}
\end{figure}

Another problem in the process of the mesh refinement is how to
treat the refinement boundary well to avoid possible numerical
noise. Here we follow the guard cell scheme described in
\cite{imbiriba04}. This method has three steps, shown
in Fig.~\ref{guard}. In the first step, interior fine grid cells
are used to fill the interior grid cells of the next lower level.
This restriction operation is depicted for the case of two spatial
dimensions in the left panel of Fig.~\ref{guard}. The restriction
is basically a three-dimensional interpolation in this work, and
is accurate to fourth-order in the grid spacing to match the
accuracy of the finite-differencing scheme. As in the left panel
of Fig.~\ref{guard}, the values of the coarser grid cells (red
circle with black rim) located within the finer grid cells (green
diamonds) are filled with the finer values by quartic interpolations. The
stencil of the finer cells needs to take care to ensure that
only interior fine grid points, and no fine grid guard cells, are
used in this first step. Secondly, the fine grid guard cells (red
diamonds with black rims) are filled by prolongation from the grid
of the next lower level. Before the prolongation, the coarser grid
updates its own cells (green circles with black rim in the right
panel of Fig.~\ref{guard}) from the finer grids in the first step.
The stencil used in the prolongation operation is shown in the
right panel of Fig.~\ref{guard}. The prolongation is a
three-dimensional interpolation, and is also accurate to
fourth-order, like the restriction in the first step. In this
case, the coarser grid stencil includes two layers of guard cells
(green circles), as well as its updated interior grid points
(green circles with black rims). In the last step, the coarser
guard cells close to the interface (bold line) are filled by using
derivative matching, the difference between the finer cell and the
neighbor finer guard cell across the interface matches the
difference between the coarser cell and the neighbor coarser cell
across the interface \footnote{Sometimes we find that it is less
accurate if the last step is fulfilled, especially in the
higher-order finite-differencing scheme. We then simply neglect
the last step in the interface treatment for better accuracy.}.
%%%%%%%%%%%%%%%%%%%%%%%%%%%%%%%%%%%%%%%%%%%%%%%%%%%%%%%%%s
\section{Initial Data for punctured black holes}\label{seciv}
%%%%%%%%%%%%%%%%%%%%%%%%%%%%%%%%%%%%%%%%%%%%%%%%%%%%%%%%%%
For the initial data of punctured black holes, we consider the
Bowen-York type initial data, in which the maximal slicing and the
conformal flat form are adopted \cite{brandt97}. Let $\psi$ be the
conformal factor, $\psi\equiv e^\phi$. The conformal extrinsic
curvature reads
\begin{eqnarray}
\tilde{A}_{ij}&=&\psi^{-6}\hat{K}_{ij}=\frac{3}{2}
\sum_I\frac{\psi^{-6}}{r^{2}_I}[2P^I_{(i}n^I_{j)}\nonumber\\
&&\quad-(f_{ij}-n^I_{i}n^I_{j})P_I^{k}n^I_{k}
+\frac{4}{r_I}n^I_{(i}\epsilon_{j)k\ell}S_I^{k}n_I^{\ell}],\label{YorkA}
\end{eqnarray}
where $f_{ij}$ is the three flat metric; $P_I^i$ and $S_I^i$ are
constant vectors, standing for the linear momentum and spin momentum
of the $I$-th black hole respectively; $n_I^i$ is the radial
normal vector with respect to $f_{ij}$, which points from the
position of the $I$-th black hole to the space point. In the
puncture method described in \cite{brandt97},
$\psi=1+\sum\frac{m_I}{2r_I}+u$ with mass parameter $m_I$ for the
$I$-th black hole, and $u$ is determined by
\begin{equation}
(\partial^2_x+\partial^2_y+\partial^2_z)u=-\frac{1}{8}\hat{K}^{ij}\hat{K}_{ij}
(1+\sum_I\frac{m_I}{2r_I}+u)^{-7}.\label{by_eq}
\end{equation}
If all black holes are at rest and spinless, (\ref{YorkA}) implies
$\hat{K}_{ij}=0$, so $u=0$. In the head-on collision case, we will
implement this kind of initial data with $m_1=m_2=0.5$. For a
single black hole, when the linear momentum and the spin are
small, we can solve the above equation approximately as (see
Appendix \ref{spinID} for more detail) \cite{small_p}
\begin{eqnarray}
u=\frac{{\vec
P}^2}{m^2}[u_1+u_2(3\mu_P^2-1)]&+&6\frac{u_3}{m^4}{\vec S}^2
(1+\mu_S^2)\nonumber\\
&+&\frac{u_4}{m^3}{\vec P}\times{\vec S}\cdot{\vec{n}},\label{IDsoln}
\end{eqnarray}
with
\begin{eqnarray}
u_1&=&\frac{5\ell}{8}(1-2\ell+2\ell^2-\ell^3+\frac{1}{5}\ell^4),\nonumber\\
u_2&=&\frac{1}{40b^2}(15+117\ell-79\ell^2+43\ell^3\nonumber\\
&&\qquad\qquad-14\ell^4+2\ell^5+84\ln\ell/b),\\
u_3&=&\frac{\ell}{20}(1+\ell+\ell^2-4\ell^3+2\ell^4),\nonumber\\
u_4&=&\frac{\ell^2}{10}(10-25\ell+21\ell^2-6\ell^3),\nonumber
\end{eqnarray}
where $b=2r/m$, $\ell=1/(1+b)$, $\mu_P={\vec{P}}\cdot{\vec{n}}/P$
and $\mu_S={\vec{S}}\cdot{\vec{n}}/S$. For the approximate
solution (\ref{IDsoln}) of a moving black hole with spin, the ADM
mass, the linear momentum, and the angular momentum are
\begin{eqnarray}
M_{\rm ADM}&=&m+\frac{5}{8}\frac{{\vec P}^2}{m}
+\frac{2}{5}\frac{{\vec S}^2}{m^3},\label{admmass}\\
{\vec P}_{\rm ADM}&=&{\vec P},\\
{\vec S}_{\rm ADM}&=&{\vec S}.
\end{eqnarray}

On the other hand, when $\vec{P}=0$ while $\vec{S}$ is very large,
the conformal factor can be approximated as (see Appendix
\ref{spinID} for more detail) \cite{lovelace08}
\begin{eqnarray}
\psi=\frac{(6S^2)^{1/8}}{\sqrt{r}}.\label{high_spin_solution}
\end{eqnarray}

In this paper we will fix $m=1$ for all single black hole
simulations.
%%%%%%%%%%%%%%%%%%%%%%%%%%%%%%%%%%%%%%%%%%%%%%%%%%%%%%%%%%
\section{Numerical results}\label{secv}
%%%%%%%%%%%%%%%%%%%%%%%%%%%%%%%%%%%%%%%%%%%%%%%%%%%%%%%%%%
In this section we report the numerical results for: (1) a single
moving black hole without and with spin: The moving action and the
spinning action of a single black hole are fundamental elements
for BBH simulations. As our code aims at simulating BBH
coalescence, evolving a single moving and spinning black hole
becomes an essential test. In addition, since the gauge choice is
critical for the moving puncture method, we would like to study if
there are any other gauge conditions which can also support the
moving puncture technique, besides the known ones. Our main
achievement in this part of the work is that we discover one new
gauge condition besides the known ones which can support moving
puncture black hole simulations. The results on the gauge
condition tests are listed in Table \ref{shifttype}, where Gauge
VII is the aforementioned new set of gauge conditions. The
successes of the gauge usage in other groups' work, e.g.,
\cite{moving_punctur1,moving_punctur2,brugmann08,vaishnav07,SpeU07,meter06},
are also reconfirmed in Table \ref{shifttype}. (2) the head-on
collisions of BBHs: The head-on collision of a BBH system is the
simplest dynamical spacetime in which a complete gravitational
waveform of the merger of two black holes could be produced.
Therefore, we use this scenario to examine the performance of the
code. Besides, in order to go beyond the cases of a head-on
collision in vacuum, the case of a head-on collision perturbed by
a massless scalar field is also studied. With such kinds of cases
we try to understand qualitatively the effect of the existence of
neutral matter on the collision of a BBH, especially on its
gravitational radiation. It is shown that the waveform could be
affected significantly by the scalar field.

%%%%%%%%%%%%%%%%%%%%%%%%%%%%%%%%%%%%%%%%%%%%%%%%%%%%%%%%%%
\subsection{Static Black Hole}\label{secva}
%%%%%%%%%%%%%%%%%%%%%%%%%%%%%%%%%%%%%%%%%%%%%%%%%%%%%%%%%%
\begin{figure}[t]
\begin{tabular}{c}
\includegraphics[width=0.47\textwidth]{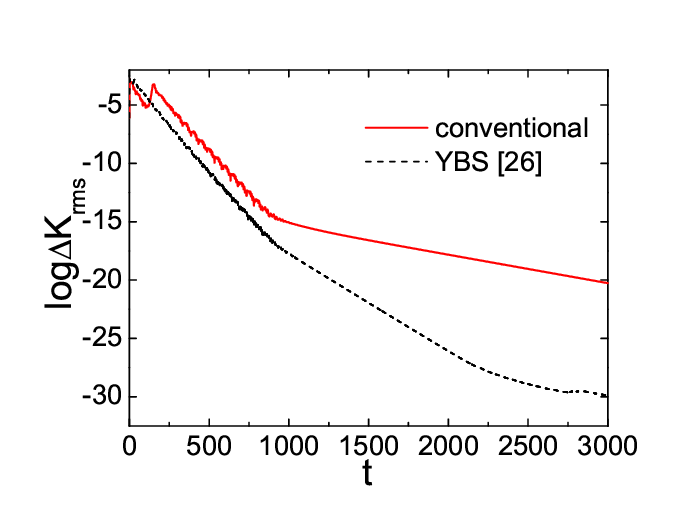}
\end{tabular}
\caption{The root mean square of the change in the trace of
extrinsic curvature between consecutive time steps as a function of
time in the static case with equatorial symmetry. The solid (red)
line is the result with the conventional setting. The
dashed line is the result with the modifications
suggested in \cite{YHBS02}. This shows that the both settings give
stable and convergent results.}
\label{fig0}
\end{figure}

Although the BSSN formulation with the ``1+log'' lapse condition
and the $\Gamma$-driver shift condition has been shown to be
well-posed and hyperbolic \cite{gundlach06a,BeHS04}, it is still
useful to confirm the stability and convergence of the formulation
and the applied modifications before we move forwards to the
moving/spinning black hole cases in the following subsections.
Therefore, for the equation adjustments, we enforce the
constraints (\ref{detg1})--(\ref{Gofg}) by using
Eqs.~(\ref{gamave}) and (\ref{Aave}), and the substitution of the
conformal connection ${\tilde\Gamma}^i$ with
$-\tilde\gamma^{ij}{}_{,j}$, as described in Sec. \ref{seciib}.
For the gauge condition, Eqs.~(\ref{lapse_eq})--(\ref{b_eq}) are
applied with $f(\alpha)=\alpha$ and the parameter choice
$\lambda_3=0$, $\lambda_1=\lambda_2=\lambda_4=1$, which is close
to Gauge VII in Table \ref{table1}, i.e., the newly viable gauge
condition (see Section \ref{secvb}). The grid width $h=0.2$ and
the outer boundary is $\pm16$, $\pm16$, $16$, respectively,
assuming equatorial symmetry. In this simple case we only consider
a unigrid for the computational domain.

\begin{table}[htbp]
\caption{The gauge choices tested in this work correspond to
Eqs.~(\ref{lapse_eq}), (\ref{beta_eq}) and (\ref{b_eq}); ``$\vee$"
stands for the gauge condition with the corresponding advection
term, while ``$\times$" stands for without.} \label{shifttype}
\begin{ruledtabular}
\begin{tabular}{cccccc}
Gauge No.&$\beta^i\partial_i\alpha$&$\beta^j\partial_j\beta^i$&
$\beta^j\partial_jB^i$&$\beta^j\partial_j\tilde{\Gamma}^i$&Tests\\
\hline I   &$\times$&$\times$&$\times$&$\times$&FAIL\\
\hline II  &$\vee$  &$\times$&$\times$&$\times$&PASS\\
\hline III &$\vee$  &$\vee$  &$\times$&$\times$&FAIL\\
\hline IV  &$\vee$  &$\times$&$\vee$  &$\times$&FAIL\\
\hline V   &$\vee$  &$\times$&$\times$&$\vee$  &FAIL\\
\hline VI  &$\vee$  &$\vee$  &$\vee$  &$\times$&FAIL\\
\hline VII &$\vee$  &$\vee$  &$\times$&$\vee$  &PASS\\
\hline VIII&$\vee$  &$\times$&$\vee$  &$\vee$  &FAIL\\
\hline IX  &$\vee$  &$\vee$  &$\vee$  &$\vee$  &PASS
\label{table1}
\end{tabular}
\end{ruledtabular}
\end{table}

The result for such a conventional setting is shown in
Fig.~\ref{fig0}. This figure shows a log plot for the root mean
square (r.m.s.) of the changes in the trace of extrinsic curvature
$K$ (the solid red line) between consecutive time steps. In the
plot the curve of the change in $K$ rises around $t=200$ during
the period of settlement. The change in $K$ decreases
exponentially afterwards, without a sign of rise to the end of the
run. This indicates that the conventional settings give a stable
evolution for a single static black hole. The result is also
consistent with the analytic understanding of the BSSN formulation
and of the gauge choice.

Meanwhile, some modifications \cite{YHBS02}, especially the
enforcement of the constraints (\ref{detg1})--(\ref{Gofg}), have
been shown to be at least as good as the conventional
ones in the numerical result \cite{SpeU07}. We are therefore
interested in understanding the performance of the code with the
modifications, and the comparison between these two sets.

Briefly, the modifications in \cite{YHBS02} are summarized as
follows: Instead of treating all components of
${\tilde\gamma}_{ij}$ equally, only five of the six components of
${\tilde\gamma}_{ij}$ are evolved dynamically, and the
$zz$-component is computed using Eq.~(\ref{detg1}),
\begin{equation}
{\tilde\gamma}_{zz}=1+\frac{{\tilde\gamma}_{yy}{\tilde\gamma}_{xz}^2
-{\tilde\gamma}_{xy}{\tilde\gamma}_{yz}{\tilde\gamma}_{xz}
+{\tilde\gamma}_{xx}{\tilde\gamma}_{yz}^2}{{\tilde\gamma}_{xx}
{\tilde\gamma}_{yy}-{\tilde\gamma}_{xy}^2}.
\end{equation}
Similarly, ${\tilde A}_{zz}$ is determined from the other five
components of ${\tilde A}_{ij}$ using Eq.~(\ref{trA0}),
\begin{equation}
{\tilde A}_{zz}=-\frac{{\tilde A}_x{}^x+{\tilde A}_y{}^y
+{\tilde A}_{xz}{\tilde\gamma}^{xz}+{\tilde A}_{yz}{\tilde\gamma}^{yz}}
{{\tilde\gamma}^{zz}}.
\end{equation}
Instead of substituting for the undifferentiated conformal connection
${\tilde\Gamma}^i$ by $-\tilde\gamma^{ij}{}_{,j}$ according to the constraint
(\ref{Gofg}), the constraint is added to the evolution equation (\ref{dtGamma})
of ${\tilde\Gamma}^i$ via
\begin{equation}
\partial_t{\tilde\Gamma}^i=\mbox{rhs of }(\ref{dtGamma})-(\xi+2/3)
({\tilde\Gamma}^i+\tilde\gamma^{ij}{}_{,j})\beta^k{}_{,k},
\end{equation}
where $\xi$ is usually chosen to be $2/3$.

With otherwise the same settings as in the conventional ones, the
result with the modifications in \cite{YHBS02} is also indicated
by the dashed line in Fig.~\ref{fig0}. Without too long a
settlement period, the change in $K$ decreases exponentially to
the end of the run. A comparison of these two sets of
modifications shows that they are equally good in converging the
evolution to a numerically stable state, although the
modifications in \cite{YHBS02} give a better settlement at the
early stage of the evolution. This indicates that there is still
room for modifying the BSSN formulation to achieve better stability and
convergence. The major purpose of this work is to build a reliable code
for BBH simulations, therefore we will stick to the conventional
modifications in the rest of this paper, although the
modifications in \cite{YHBS02} might show some subtle advantage in
stabilization over the conventional ones.
%%%%%%%%%%%%%%%%%%%%%%%%%%%%%%%%%%%%%%%%%%%%%%%%%%%%%%%%%%
\subsection{Moving Black Hole without Spin}\label{secvb}
%%%%%%%%%%%%%%%%%%%%%%%%%%%%%%%%%%%%%%%%%%%%%%%%%%%%%%%%%%
\begin{figure*}[thbp]
\begin{tabular}{c}
\includegraphics[width=0.93\textwidth]{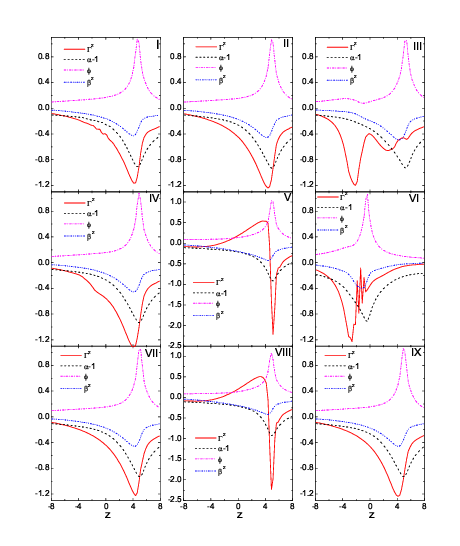}
\end{tabular}
\caption{The gauge tests for a moving black hole without spin
(velocity $v\approx0.615$). The profiles of several dynamical
variables at $t=30$ are shown, except for case VI, for which the
time is $t=12$. The horizontal axis is the $z$-axis, the moving
direction of the black hole. The vertical axis is the
corresponding value for different variables: the solid (red) line
is $\tilde{\Gamma}^z$; the dashed line is $\alpha-1$; the
(magenta) dot-dashed line is $\phi$; the (blue) dot-dot-dashed
line is $\beta^z$. The different panels correspond to the
different cases in which the gauge number is marked on the
upper-right corner of the panel and they are also listed in Table
\ref{table1}. For Gauges III, V, and VIII, the results show some
ill behavior explicitly while Gauges I and IV have tails of noise
behind the black hole. The rest of the three gauge choices II,
VII, and IX give almost the same well-behaved result. The results
of Gauges II and IX are consistent with other research groups'
work. Gauge VII is found to work well with the moving
puncture method in this work.} \label{fig1}
\end{figure*}

We now study the cases of a moving black hole without spin, which
are similar to the ones in \cite{meter06}. Three levels of grids
are used in this and the next two subsections. The outer boundary
of the coarsest level is set at $\pm16$, and the boundaries of the
finer levels are located at $\pm8$ and $\pm4$, respectively. The
gridwidth for the highest level is $1/8$. The black hole is
located at $(0,0,-3)$ initially with the linear momentum vector
$\vec P=(0,0,1)$. The gauge choices tested in this work are listed
in Table \ref{shifttype}. Gauges I, II, IV, V, VI and IX (Figs.~3,
5, 7, 8, 6, 10 in \cite{meter06}, respectively) have been tested
in \cite{meter06}. Differing from the numerical initial data used
in \cite{meter06}, the approximate analytic initial data described
in Sec.~\ref{seciv} is used in the tests. Nevertheless, our
results are consistent with the results in \cite{meter06}.

We summarize our results obtained for the nine tested gauges in
Fig.~\ref{fig1}. In each panel of Fig.~\ref{fig1}, the number on
the upper-right corner indicates the case with the same gauge
number in Table \ref{table1}, and the result of that case is
plotted in the panel. In each case, the black hole moves along the
$z$-axis, and the conformal connection $z$-component
${\tilde\Gamma}^z$ (the red solid line), the lapse function
$\alpha$ (the dashed line), the conformal exponent $\phi$ (the
magenta dot-dashed line), and the shift vector $z$-component
$\beta^z$ (the blue dot-dot-dashed line) are chosen as the
monitors for the stability of each run. The profiles of these
variables are recorded at time $t=30$ in each panel, except for
Panel VI, where the recorded time is $t=12$. In Panel I, it can be
seen that $\alpha$, $\phi$, and $\beta^z$ all behave well, but
${\tilde\Gamma}^z$ has some tail of ripple behind the black hole.
The ripple implies a rising instability in the evolution. It is
known that adding the $\alpha$ advection term in
Eq.~(\ref{lapse_eq}), i.e., $\lambda_1=1$, suppresses this
unstable mode. We verify it by comparing the profiles of the
variables in Panel I and II. In Panel II, all variables behave
well and there is no ripple of noise for the curve of
${\tilde\Gamma}^z$. This result is consistent with the one in
\cite{lousto08}. We then set $\lambda_1=1$ in Eq.~(\ref{lapse_eq})
in the following cases, since it is necessary for the stability of
a moving black hole.

%The instability always relates to the grid points with the extreme
%value of $\tilde{\Gamma}^z$. Because this point corresponds to the
%position of moving black hole, while the role of $\tilde{\Gamma}$
%in the stability of BSSN formalism is key important. In addition,
%when the gauge choice results in stable simulation, the profile of
%conformal connection always follows the shape of the profile of
%shift vector. This is because the gauge condition of moving
%puncture depends on the synchronization of shift and conformal
%connection.

In Panel III, it is shown that the addition of the advection term
of $\vec\beta$ in Eq.~(\ref{beta_eq}), i.e.,
$\lambda_1=\lambda_2=1$, will result in a big bump behind the
black hole. This instability is somehow strong, such that it
causes the ill-behavior of $\alpha$ and $\phi$. Adding the
advection term of $\vec B$ in Eq.~(\ref{b_eq}), i.e.,
$\lambda_1=\lambda_3=1$, will bring in a small tail, as seen in
Panel IV, although such an instability does not seemingly affect
$\alpha$, $\phi$, and $\beta^z$, for they behave well in the plot.
The subtraction of the advection term of $\tilde{\Gamma}^i$ in
Eq.~(\ref{b_eq}), i.e., $\lambda_1=\lambda_4=1$, will result in a
``distorted'' profile for $\tilde{\Gamma}^z$ in Panel V. (A gauge
choice close to Gauge V has been used to simulate the inspiral of
a BBH system with a small initial separation in
\cite{moving_punctur2}.) From the above three cases, we see that
solely adding the advection term of $\vec\beta$ or $\vec B$, or
subtracting the advection term of $\tilde{\Gamma}^i$ will bring in
instability in general. This understanding leads us to try
combinations of these three cases.

\begin{figure*}[ht]
\includegraphics[totalheight=0.5\textheight,width=\textwidth]{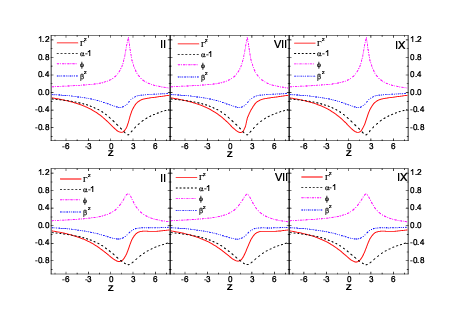}
\caption{The gauge tests for a moving black hole with spin
parallel (upper panel) and perpendicular (lower panel) to the
moving direction (velocity $v\approx0.5$, the specific angular
momentum $a\approx0.5102$). The profiles of several dynamical
variables at $t=30$ are shown as in Fig.~\ref{fig1}. From left to
right, the figures correspond to the results with Gauges II, VII
and IX listed in Table \ref{table1}, respectively. These three
``good'' gauges give almost the same results.} \label{fig2}
\end{figure*}

Panel VI shows that the combination of the $\vec\beta$ and the
$\vec B$ advection term additions, i.e.,
$\lambda_1=\lambda_2=\lambda_3=1$, results in a profile with high
frequency noise for $\tilde{\Gamma}^z$ around the black hole and
makes an even stronger instability. The code crashes soon after
time $t=12$ when the black hole has only moved slightly. However, in
Panel VII, it is shown that the combination of the $\vec\beta$
advection term addition and $\tilde{\Gamma}^i$ advection term
subtraction, i.e., $\lambda_1=\lambda_2=\lambda_4=1$, can suppress
the unstable modes introduced by the sole usage of these two terms
(see Gauges III and V). We can see that the curve profiles of all
the variables in Panel VII look almost the same as those in Panel
II. It is interesting to note that, according to our literature
search, Gauge VII has never been previously used in any BBH
simulations. Therefore, it deserves further study. In Panel VIII,
we consider the combination of adding the $\vec B$ advection term
and subtracting the $\tilde{\Gamma}^i$ advection term, i.e.,
$\lambda_1=\lambda_3=\lambda_4=1$. The performance in Panel VIII
shows a set of curve profiles which are similar to those obtained
with Gauge V. Finally, we consider in Panel IX the combination of
all three advection terms in the shift equation, i.e.,
$\lambda_1=\lambda_2=\lambda_3=\lambda_4=1$. The combination of
the BSSN equations with Gauge IX has been proven to be strongly
hyperbolic in the sense of first-order in time, second-order in
space systems \cite{gundlach06a}, and thus yields a well-posed
initial-value problem. Panel IX shows that all the variables
behave very well and smoothly. The curve profiles in Panels II,
VII, and IX look almost the same.

The linearized analysis of the BSSN formulation with the gauge
conditions in \cite{meter06} shows that both Gauges II and VII
have zero-speed modes. However, from the results described in this
section, we can not distinguish the difference between Gauges II,
VII and IX. Furthermore, Fig.~5 of \cite{meter06}, which
corresponds to gauge II, shows no zero-speed modes at all. We
conjecture that the nonlinearity of the full theory could
eliminate the zero-speed modes for Gauges II and VII in general.
 In fact, Gauge II \cite{campanelli07c,lousto08}, as
well as Gauge IX \cite{brugmann08,vaishnav07,SpeU07}, has been
successfully used in the simulations of black hole evolution. And
Gauge VII is as good as Gauges II and IX for black hole
simulations, at least in the cases tested in  this work.
Therefore, we can also expect that Gauge VII is very likely to be
viable in the generic cases of BBH evolution.

Someone may raise the doubt that the moving velocity is not large
enough to excite the zero speed mode for the newly found gauge
condition here. This is not the case in fact, at least for the
approximate initial data. Considering the ADM mass
(\ref{admmass}), we have the maximal moving velocity
$\vec{v}=\vec{P}_{\rm ADM}/M_{\rm ADM}$ for a black hole without
spin when $P=2\sqrt{10}/5$. And the moving velocity
$(v\approx0.615)$ for the above tested moving black hole with the
linear momentum $P=1$ almost equals this maximal velocity
($v\approx0.63$). We have tested this maximal moving velocity
also. The result gives the same conclusion mentioned above.
Someone may also raise the doubt that the zero speed modes for Gauge
VII mentioned in \cite{meter06} might not be excited in this
tested scenario. Therefore, we test it further, with both the case
of a moving black hole with spin and the case of a high-spin black
hole, in the following subsections. Our results will show that
Gauge VII, as well as Gauges II and IX, passes these two tests.
%%%%%%%%%%%%%%%%%%%%%%%%%%%%%%%%%%%%%%%%%%%%%%%%%%%%%%%%%%
\subsection{Moving Black Hole with Spin}\label{secvc}
%%%%%%%%%%%%%%%%%%%%%%%%%%%%%%%%%%%%%%%%%%%%%%%%%%%%%%%%%%
\begin{figure*}[ht]
\includegraphics[totalheight=0.5\textheight,width=\textwidth]{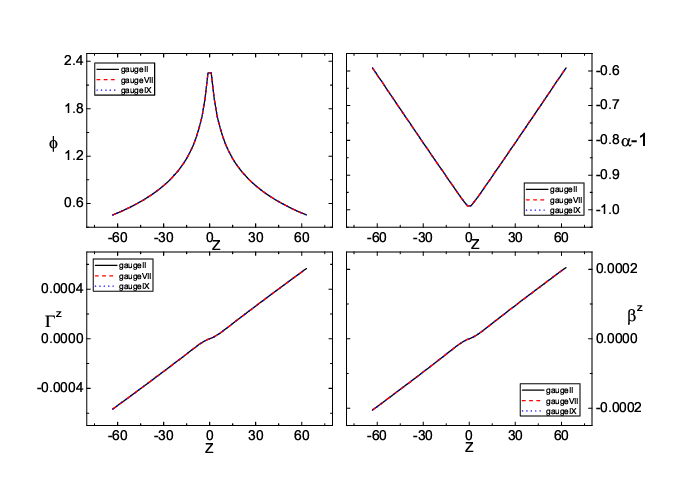}
\caption{The gauge tests for a single rapid-rotating black hole
(spin parameter $a\approx0.9$). The comparison of profiles of the
three ``good'' gauges at $t=30$ is shown. From top to bottom, and
from left to right, the panels show the conformal factor $\phi$,
the lapse $\alpha-1$, the conformal connection $\tilde{\Gamma}^z$
and the shift $\beta^z$, respectively. These three gauge choices
give almost the same results.} \label{fig3}
\end{figure*}

In this subsection, we investigate the effect of spin on a moving
black hole. Firstly, we set the spin direction of the black hole
to be parallel to its moving direction, specifically, $\vec
P=(0,0,1)$ and $\vec S=(0,0,1)$. Therefore, the only difference
between the setting here and in the previous subsection is that, in
this subsection, the black hole has a spin of amplitude $1$ which
is along the moving direction. Here we only test the ``good''
gauge choices, i.e., Gauges II, VII and IX. The curve profiles at
$t=30$ are presented in the upper panels of Fig.~\ref{fig2}. We
find that, with any of these three gauges, the code can stably
simulate a spinning black hole. The curves in the three upper
panels look the same, not showing a sign of instability. Compared
with Fig.~\ref{fig1}, we find that the black holes in this case
move slower ($v\approx 0.5$) than the spinless ones ($v\approx
0.615$) with the same gauge choices described in the previous
subsection. This is consistent with the theoretical prediction:
From Eq.~(\ref{admmass}) we can see that the moving black hole
with a nonvanishing spin will have a larger ADM mass than the
spinless one due to the rotational energy. For the moving
velocity being $\vec{v}=\vec{P}_{\rm ADM}/M_{\rm ADM}$,
a spinning black hole will move slower than a spinless one
with the same linear momentum. Our conclusion in this test is
that Gauges II, VII and IX all handle this scenario well, since
these three gauges give almost the same result.

Secondly, we set the spin direction of the black hole to be
perpendicular to the moving direction, specifically, $\vec
P=(0,0,1)$ and $\vec S=(1,0,0)$. Similarly, we only test Gauges
II, VII and IX. The curve profiles at $t=30$ are presented in the
lower panels of Fig.~\ref{fig2}. The results are similar to those
obtained in the parallel-spin case. From Eq.~(\ref{admmass}), the
moving velocity of the black hole for this case is the same as the
velocity in the parallel-spin case, i.e., $v\approx 0.5$. We can
read this from the results presented in Fig.~\ref{fig2}. In this
case, the peak amplitude of the variable $\phi$ becomes smaller
than in the parallel-spin case. However, this difference basically
comes from the spin orientation; it does not bring any instability
to the results with the three ``good'' gauges II, VII and IX.
These three ``good'' gauges give essentially the same stable behavior.

Similar to the moving velocity being bounded, the magnitude of the
specific angular momentum, $\vec a=\vec{S}/M^2_{\rm ADM}$, has a
maximal value for the approximate initial data (\ref{IDsoln}). For
$S=\sqrt{5/6}$, the black hole has a maximal magnitude
$a=3\sqrt{30}/32\approx0.5135$. The magnitude of $\vec a$ in both
the parallel-spin and perpendicular-spin cases is
$a\approx0.5102$. Therefore, the magnitude of $\vec a$ in these
two cases is very close to the maximal magnitude of the specific
angular momentum that the initial data can have. In the next
subsection we would like to test these three gauge choices with an
even higher spin magnitude of $\vec a$.
%%%%%%%%%%%%%%%%%%%%%%%%%%%%%%%%%%%%%%%%%%%%%%%%%%%%%%%%%%
\subsection{Rapidly-rotating black hole}
%%%%%%%%%%%%%%%%%%%%%%%%%%%%%%%%%%%%%%%%%%%%%%%%%%%%%%%%%%
We have tested Gauge VII in the spinless and the spinning moving
black hole cases. We find that, with this gauge choice, the
code can simulate these dynamical spacetimes well. Since
experience tells us that a higher spin is more likely to trigger a
quicker instability in the numerical simulations, one might doubt
whether the newly found gauge condition can handle a high-spin
case well. In this subsection, we test the three ``good'' gauge
choices with a rapidly-rotating black hole. The initial data used
is another approximate analytic solution for the Bowen-York
initial data. We describe the details in Appendix \ref{spinID}.
This initial data is similar to that used in Sec.~\ref{secvc},
except that the conformal factor $\psi$ is given by
Eq.~(\ref{high_spin_solution}) rather than Eq.~(\ref{IDsoln})
\cite{dain08,lovelace08} and $\vec P=0$.

Here we set the angular momentum vector to be $\vec
S=(0,0,10000)$, which results in the specific angular momentum
$a\approx0.9$ \cite{lovelace08}. In this case, we again only care
about the corresponding behavior of the runs with the three
``good'' gauge choices. The outer boundary is set at $r=128$, and
six levels of grids for FMR are used in the runs. The curve
profiles at $t=30$ are presented in Fig.~\ref{fig3}. These three
``good'' gauge choices can all give a stable and accurate
simulation for this highly spinning single black hole case.
Looking at the plots in this figure, it is difficult to
distinguish between the results obtained with these three gauges.
The curves overlap each other well. The profiles of
$\tilde{\Gamma}^z$ and $\beta^z$ are consistent with the
experience that they almost give the same shape. Due to the limits
on computational resources, the runs are stopped at $t=30$. This
might raise the doubt whether the runtime is long enough to excite
an unstable mode. Since the spin of the black hole in this case is
close to the maximal one ($a_{\rm max}\approx 0.9282$) that a
punctured black hole can have in a Bowen-York type initial data
\cite{lovelace08}, we expect that some difference between these
runs, caused from instability, will appear in this period of time.
However, we find that the results with these three gauge choices
are almost the same. This should indicate that Gauge VII is as
good as the other two in this case, unless all three of these
gauge choices cause the same instability, which is highly
unlikely. Meanwhile, the results in Sec.~\ref{secva} can be
regarded as complementary to the one in this subsection (about the
the long-term stability issue of Gauge VII). Therefore we conclude
that the new gauge choice, Gauge VII, survives in this high-spin
test.
%%%%%%%%%%%%%%%%%%%%%%%%%%%%%%%%%%%%%%%%%%%%%%%%%%%%%%%%%%
\subsection{Head-on Collision of two equal mass black holes}
%%%%%%%%%%%%%%%%%%%%%%%%%%%%%%%%%%%%%%%%%%%%%%%%%%%%%%%%%%
In the previous subsections, we have presented the numerical
simulations for a single black hole with/without spin with the
different gauge choices. In order to test our code as well as the
three gauge choices discussed above further, we present
in this and the next subsections the
numerical results about the head-on collisions of a BBH system,
which is the simplest case for BBHs.
We use time-symmetric initial data for the two black
holes, i.e., the so-called Brill-Lindquist initial data
\cite{brill64}. Specifically, the initial data takes the form
\begin{eqnarray}
e^\phi&=&1+\frac{m_1}{2|\vec{r}-\vec{c}_1|}
          +\frac{m_2}{2|\vec{r}-\vec{c}_2|},\\
\tilde{\gamma}_{ij}&=&f_{ij},\qquad K=0,\qquad\tilde{A}_{ij}=0.
\end{eqnarray}
where $m_1$ and $m_2$ are the mass parameters for the two
black holes, $\vec{c}_1$ and $\vec{c}_2$ are the positions of the
black holes, and $f_{ij}$ stands for the flat three-metric. In
this case, we set the mass parameters to be $m_1=m_2=0.5$ and the
positions of the two black holes at $(0,0,\pm1.1515)$,
respectively. This value has been used in
\cite{cook94,SpeU07,alcubierre03} and corresponds to an initial
separation of the black holes equal to that of an approximate ISCO
configuration. In other words, our initial data corresponds to two
identical black holes which have no spin and no linear momentum,
attracting each other from rest at the ISCO.

\begin{figure}[t]
\includegraphics[width=0.48\textwidth]{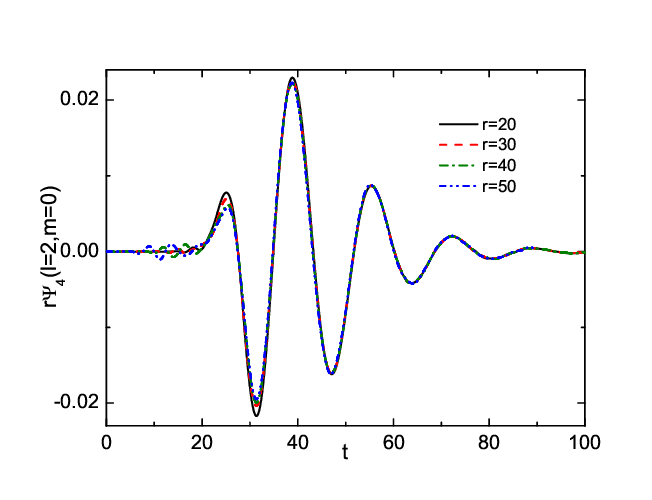}
\caption{The ($\ell$=2, $m$=0) mode of $r\Psi_4$ extracted from a head-on
collision of the Brill-Lindquist initial data starting from the approximate
ISCO separation $d=2.303$ at four different radii $r$=20, 30, 40 and 50.
Considering the velocity of the gravitational wave, the time delay has been
shifted in the plot.}
\label{waveR_chi_headon}
\end{figure}

In this and the next subsections, the computational domain is
$\pm 64\times\pm 64\times 64$ and $64\times64\times32$
grid points are used in every level.
To reduce the computational load equatorial symmetry is assumed.
Six levels of grids are used for the mesh refinement. The
refinement boundaries are placed at 32, 16, 8, 4, and 2. As
mentioned in Sec.\ref{seciib}, the $\chi$-version of the evolution
equations are also available in our code. In this and the next
subsections, we test both the $\phi$-version and the
$\chi$-version of the code in the head-on collision cases. The
results obtained from these two versions are consistent in the
scenarios \footnote{In this work, we mainly stick to the
$\phi$-version of the code. However, in our experience, the
$\chi$-version of the code sometimes gives better stability and
convergence, although it also suffers from a possible problem
during the evolution, due to $\chi$'s turning negative near the
punctures.}.

\begin{figure}[t]
\includegraphics[width=0.48\textwidth]{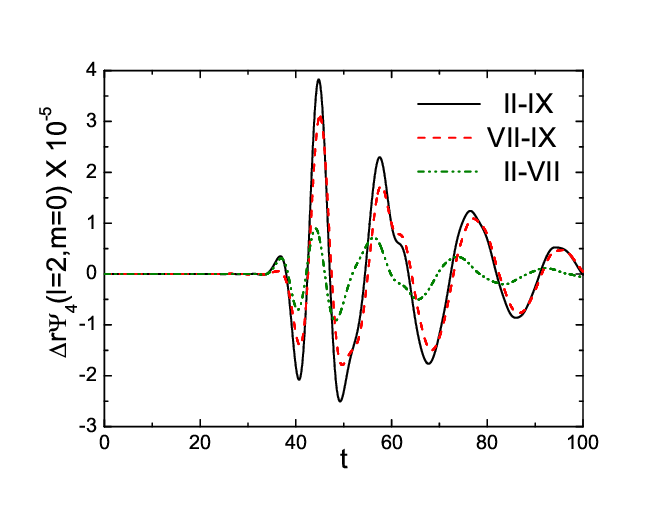}
\caption{The differences between the ($\ell$=2,$m$=0) mode of the waveform
$r\Psi_4$
for different gauge choices extracted at $r=30$.
The solid line is the difference between Gauges II and IX,
the (red) dashed line is the difference between Gauges VII and IX,
and the (olive) dot-dot-dashed line is the difference between Gauges II and VII.
The largest difference is roughly $0.2\%$ of the amplitude of $r\Psi_4$.}
\label{waveR_chi_gauge}
\end{figure}

Although conceptually simple, the head-on collision of a BBH
system is still a highly dynamical process in spacetime. During
the collision, the system will emit a complete gravitational
waveform from the merger of two black holes. To quantify the
gravitational waveform, we use the Newman-Penrose scalar
$\Psi_{4}$. The method of computing this quantity is described in
Appendix \ref{appendix_psi4}. Due to the symmetric property of
this BBH system, the gravitational wave has only the
($\ell$=2,$m$=0) mode. First, we consider Gauge IX, which has been
widely adopted in simulations of BBH systems. We extract the
waveform at $r$=20, 30, 40, and 50. Since the leading order of
$\Psi_4$ is $1/r$ asymptotically, $r\Psi_4$ should be independent
of the extraction position, except for the time delay due to the
gravitational wave propagation. We consider the velocity of the
gravitational wave to be the speed of light and thus subtract the
time delay accordingly. The waveforms are plotted in
Fig.~\ref{waveR_chi_headon}. The result is quantitatively
consistent \footnote{Note the difference of the tetrad by a factor
of two between our setting and the setting in \cite{fiske05}. This
results in the difference in the magnitude of the waveform by a
factor of two.} with the result reported in \cite{fiske05,SpeU07}.
Initially there seems to be some small amplitude oscillations
before the larger oscillations for the monitors at larger radii
$r$, but not for the one at smaller $r$. This noise mainly comes
from the reduced accuracy of the evolution in the coarser grids.

We next study the effect of the gauge conditions discussed in the
previous subsections in the head-on collision. The differences of
the waveforms with the gauge choices are shown in
Fig.~\ref{waveR_chi_gauge}. As expected, the codes with these
three ``good'' gauge conditions can all evolve the head-on
collision process stably. However, the gauge conditions indeed
affect the waveform \cite{brown08}. The highest peak of the
difference between Gauges II and IX, $\Delta(r\Psi_4)_{\rm
II-IX}$, is about $4\times10^{-5}$. From the plot, we can see that
the pattern and the amplitude of $\Delta(r\Psi_4)_{\rm VII-IX}$,
the (red) dashed line, is close to $\Delta(r\Psi_4)_{\rm II-IX}$,
the solid line. The amplitude of the wave is about $0.02$, so the
relative difference is about $0.2\%$ of the amplitude of
$r\Psi_4$. But it is interesting to note that the amplitude of
$\Delta(r\Psi_4)_{\rm II-VII}$, the (green) dot-dashed line, is
smaller, only about $0.05\%$. We note that these differences are
smaller than the difference of wave form resulting from the
different initial lapse profile (for example $\alpha=1$,
$1/\psi^2$, $1/\psi^4$, etc.) which is typically $1\%$ of the
amplitude of $r\Psi_4$.

\begin{figure}[t]
\includegraphics[width=0.48\textwidth]{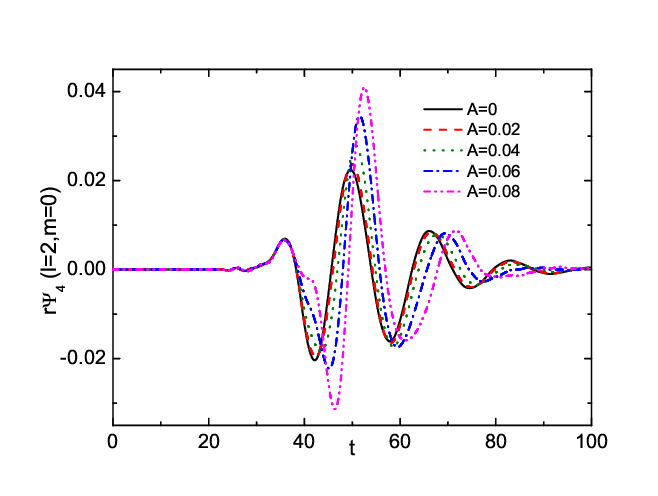}
\caption{The ($\ell$=2,$m$=0) mode of the waveform $r\Psi_4$ for a
massless scalar field perturbing the head-on collision of two
identical black holes without spin or linear momentum. $A$ is the
strength of the perturbation. $A=0.0$ means without perturbation.
The extraction radius is $r=30$. The time delay effect of the
perturbation is clear. The perturbation will efficiently amplify
the waveform.} \label{waveR_chi_scalarP}
\end{figure}
%%%%%%%%%%%%%%%%%%%%%%%%%%%%%%%%%%%%%
\subsection{Head-on Collision Perturbed by a Massless Scalar Field}
%%%%%%%%%%%%%%%%%%%%%%%%%%%%%%%%%%%%%
In the previous case, we studied the head-on collision of a BBH
system without matter. However, matter is attracted into strong
gravitational systems. Thus matter also plays an important role in
binary compact objects. The evolution of a dynamic spacetime
generally needs to include matter. On the other hand, most of the
astrophysical systems, including BBH mergers, are usually
surrounded by an accretion disk or some kind of matter. Therefore,
it is interesting to ask how this matter will affect the
gravitational wave signal, which is expected to be detected in the
near future. In order to check the consistency with the matter
coupling in the code and to investigate the effect on the
gravitational waveform by matter, we study the head-on collision
``perturbed'' by a scalar field in this subsection. Here
``perturbed'' means that the amplitude of the scalar field is
small, and we do not need to solve the constraint equation to get
exact initial data. Instead, we set the dynamical variables of
geometry to be identical to those in the previous subsection.
Furthermore, the matter part of the dynamical variables are set
independently. However when we evolve this initial data, we do not
do any approximation, that is to say we solve the fully coupled
Einstein-Klein-Golden equation numerically.

The evolution equation of a scalar field is described in Appendix
\ref{appendix_scalar}. For each simulation, we provide the
following initial data for $\Phi$ and $\partial_t\Phi$: the
profiles of $\Phi$ are a rest sphere shaped scalar field located
at the center of the two black holes
\begin{equation}
\Phi(t=0)=Ae^{-r^2}, \qquad\partial_t\Phi(t=0)=0,
\end{equation}
where $A$ is the amplitude of the scalar field. We test the scalar
field with different amplitudes to see its effect on the waveform.
The amplitudes are set to be $A$=0.02, 0.04, 0.06, and 0.08. The
results are plotted in Fig.~\ref{waveR_chi_scalarP}. It is clear
to see that the waveform is delayed with the existence of the
scalar field, and the larger the amplitude of the scalar field is,
the longer the delay time. In the meantime, the amplitude of the
waveform also becomes larger. This phenomenon can be understood as
follows: The scalar field is located at the middle of the black
hole. Some part of the scalar field will escape outwards when it
evolves. This escaping part of the scalar field will delay the
motion of the two black holes toward each other. Meanwhile, some
part of the scalar field will be absorbed into the black holes.
Thus, the black holes will become larger, and this results in a
larger amplitude of the gravitational waveform.
%%%%%%%%%%%%%%%%%%%%%%%%%%%%%%%%%%%%%%%%%
\section{summary}\label{secvi}
%%%%%%%%%%%%%%%%%%%%%%%%%%%%%%%%%%%%%%%%%
In summary, we have constructed from scratch a new numerical code based on the
BSSN formalism and the moving puncture technique. In
the code, an FMR/AMR algorithm is implemented via the GrACE
package, a fourth-order spatial finite-differencing scheme and an
iterative Crank-Nicholson scheme for time integrations are applied
in solving the Einstein equations. Some adjustments of the BSSN
formulation for the constraint equations in \cite{YHBS02} are also
examined in the work. We have compared the alternative adjustments
with the conventional ones with a static Schwarzschild black hole
and found that with both of them the black hole can be evolved
stably and accurately. We then investigated the viability of
several gauge choices through the simulation of a single moving
black hole with and without spin. In addition to obtaining results
consistent with those of other researchers, we found a new gauge
choice with which one can also simulate a moving punctured black
hole well. We next tested our code with the head-on collisions of
a BBH system in a vacuum and with the perturbation of a massless
scalar field. The gravitational waveform obtained in this code
from the collision in vacuum is quantitatively the same as that
obtained in the work by other groups. The purpose of the head-on
collision perturbed by a scalar field is to understand
qualitatively the effect of matter on the evolution of binary
black holes, as well as testing the code further. The result shows
that, with a specific configuration, the existence of a scalar
field could delay the merger of binary black holes, as people have
expected. The strength of the scalar field will significantly
affect the gravitational waveform.

The main goal of this work is the construction of a new code for
the study of numerical relativity. However, re-investigating the
conventional methods and exploring the alternatives to the methods
are also emphasized during the development of this code. As one
can see in Sec.~\ref{secva}, both adjustments for the constraint
equations give stable and convergent results. However,
Fig.~\ref{fig0} also gives the impression that the alternate
adjustment performs better than the conventional adjustment. This
simply indicates that there is still an issue of the optimal
choice of the constraint addition in the Einstein equations. For
the gauge condition, our study shows that a new gauge choice,
i.e., Gauge VII, is able to pass all of our tests. Therefore, this
gauge could be a possible choice for BBH simulations, although
more investigation is needed. The waveform obtained from the
head-on collisions shows that with the code one can simulate the
evolution of a BBH system. This enables us not only to continue
studying numerical relativity in more complicated scenarios, like
the inspiral of binary black holes, the recoil problem, and the final
spin problem, but also to verify the existing results and to go
even farther. We believe there will never be too much
double-checking of existing achievements, and any further
investigations based on them. The result of the head-on collision
perturbed by a scalar field gives us some insight into the
possible distortion of a gravitational wave, since matter is
abundant in most of the gravitating systems.

All of these results verify that the code is reliable and ready to
be engaged in the study of more realistic, astrophysical scenarios
and of numerical relativity. We plan to use this new code to work
on the inspiral of BBH systems in the very near future.
One black hole can be determined totally by only 2
parameters, i.e., mass and spin. Thus it is interesting to ask how
to determine the product black hole of a BBH system from the
information of the two initial black holes. The spin expansion
method in \cite{boyle08} shows us some hint as to how to solve
this problem, and we plan to study this problem with the new code
next.
%%%%%%%%%%%%%%%%%%%%%%%%%%%%%%%%%%%%%%%%
\section*{Acknowledgments}
%%%%%%%%%%%%%%%%%%%%%%%%%%%%%%%%%%%%%%%%
We are grateful to Dr.~Ronald Taam for helpful discussions and
encouragement and also to Dr.~Manish Parashar for offering us the
GrACE library. This work was supported in part by the National
Science Council under the grants NSC95-2112-M-006-017-MY2 and
NSC97-2112-M-006-008. Z.~Cao is supported in part by the NSFC
(Nos.~10671196 and 10731080). This work was also supported in part
by the National Center of Theoretical Sciences. We are grateful to
the National Center for High-performance Computing for the use of
computer time and facilities.

%%%%%%%%%%%%%%%%%%%%%%%%%%%%%%%%%%%%%%%%%
\appendix
%%%%%%%%%%%%%%%%%%%%%%%%%%%%%%%%%%%%%%%%%
\section{the approximate Bowen-York initial data} \label{spinID}
%%%%%%%%%%%%%%%%%%%%%%%%%%%%%%%%%%%%%%%%%
In the puncture method, the Hamiltonian constraint equation of the
Bowen-York initial data reduces to Eq.~(\ref{by_eq}). When the
linear momentum $\vec{P}$ and the spin $\vec{S}$ of a black hole
are small, we can solve the equation approximately as
\begin{equation}
(\partial^2_x+\partial^2_y+\partial^2_z)u=
-\frac{1}{8}\hat{K}^{ij}\hat{K}_{ij}(1+\frac{m}{2r})^{-7}.
\end{equation}
We chose the coordinates such that $\vec{P}$ points along the
$z$-direction and $\vec{S}$ lies on the $xz$ coordinate plane,
i.e., $\vec{S}=(S_x,0,S_z)$. In spherical coordinates, we have
\begin{eqnarray}
P^r&=&P\cos\theta,\nonumber\\
P^\theta&=&-P\sin\theta/r,\nonumber\\
P^\varphi&=&0,\nonumber\\
S^r&=&S_x\cos\varphi\sin\theta+S_z\cos\theta,\\
S^\theta&=&(S_x\cos\varphi\cos\theta-S_z\sin\theta)/r,\nonumber\\
S^\varphi&=&-S_x\sin\varphi/r\sin\theta\nonumber.
\end{eqnarray}
Note that $\epsilon_{r\theta\varphi}=r^2\sin\theta$ in spherical
coordinates. From Eqs.~(\ref{defineA}) and (\ref{YorkA}),
$\hat{K}_{ij}$ reads
\begin{eqnarray}
\hat{K}_{rr}&=&\frac{3P}{r^2}\cos\theta,\nonumber\\
\hat{K}_{r\theta}&=&-\frac{3P}{2r}\sin\theta-
\frac{3}{r^2}S_x\sin\varphi,\nonumber\\
\hat{K}_{r\varphi}&=&-\frac{3\sin\theta}{r^2}(S_x\cos\varphi\cos\theta
-S_z\sin\theta),\nonumber\\
\hat{K}_{\theta\theta}&=&-\frac{3P}{2}\cos\theta,\\
\hat{K}_{\theta\varphi}&=&0,\nonumber\\
\hat{K}_{\varphi\varphi}&=&-\frac{3P}{2}\cos\theta\sin^2\theta.\nonumber
\end{eqnarray}
Then
\begin{eqnarray}
\hat{K}^{ij}\hat{K}_{ij}
&=&\frac{9P^2}{2r^4}[\frac{5}{3}+\frac{2}{3}(3\cos^2\theta-1)]\nonumber\\
&+&\frac{9}{r^6}S_x^2[\frac{4}{3}+\frac{1}{3}(3\cos^2\theta-1)
-\sin^2\theta\cos2\varphi]\nonumber\\
&+&\frac{18S_z^2}{r^6}[\frac{2}{3}-\frac{1}{3}(3\cos^2\theta-1)]\nonumber\\
&+&\frac{18}{r^5}PS_x\sin\theta\sin\varphi
-\frac{36}{r^6}S_xS_z\cos\theta\sin\theta\sin\varphi.\nonumber\\
\label{k2}
\end{eqnarray}
By introducing the two new variables,
$\mu_P={\vec{P}}\cdot{\vec{n}}/P$ and
$\mu_S={\vec{S}}\cdot{\vec{n}}/S$, Eq.~(\ref{k2}) can be rewritten
as
\begin{eqnarray}
\hat{K}^{ij}\hat{K}_{ij}
&=&\frac{9P^2}{2r^4}[\frac{5}{3}+\frac{2}{3}(3\mu_P^2-1)]\nonumber\\
&&+\frac{18S^2}{r^6}[\frac{2}{3}-\frac{1}{3}(3\mu_S^2-1)]
+\frac{18}{r^6}(\vec{P}\times\vec{S})\cdot\vec{r}.\nonumber\\
\end{eqnarray}
Since Eq.~(\ref{by_eq}) is mainly linear with respect to  $u$, we
can solve it by dividing $\hat{K}^{ij}\hat{K}_{ij}$ into three
parts, i.e., the $P^2$ part, the $S^2$ part, and the
$\vec{P}\times\vec{S}$ part. With a variable separation method, we
can get the solution of Eq.~(\ref{by_eq}) \cite{small_p,lousto08},
i.e., Eq.~(\ref{IDsoln}).

For the scenario that $\vec{P}=0$ while the spin is large, the
solution of Eq.~(\ref{by_eq}) can be approximated with the
spherical symmetric solution \cite{lovelace08}. Then
Eq.~(\ref{by_eq}) can be smeared out as
\begin{equation}
(\partial^2_x+\partial^2_y+\partial^2_z)\psi=-\frac{3S^2}{2r^6}\psi^{-7}.
\end{equation}
In spherical coordinates, this equation reads
\begin{equation}
(\partial_r^2+\frac{2}{r}\partial_r)\psi=-\frac{3S^2}{2r^6}\psi^{-7}.
\label{high_spin}
\end{equation}
Following Dain \cite{dain08} we assume that $\psi$ behaves according to
a power-law
($\psi=Ar^\alpha$) and substitute this into Eq.~(\ref{high_spin}).
We find that the approximate solution for a highly-spinning black hole is
\begin{eqnarray}
\psi=\frac{(6S^2)^{1/8}}{\sqrt{r}}.
\end{eqnarray}
However, this approximate solution is valid only for
$m^3/S<r<S/m$, which implies that the larger the spin is, the better
the approximation is.
%%%%%%%%%%%%%%%%%%%%%%%%%%%%%%%%%%%%%
\section{wave extraction with $\Psi_4$} \label{appendix_psi4}
%%%%%%%%%%%%%%%%%%%%%%%%%%%%%%%%%%%%%
In our code we use the Newman-Penrose scalar $\Psi_4$ for extracting the
physical information from numerical simulations.
Here we follow Pretorius' work \cite{bbhsuccess} and construct the tetrad from
the timelike unit normal field as well as applying the Gram-Schmidt
orthonormalization procedure to the three-dimensional vectors
\begin{eqnarray}
\vec u&=&[x,y,z],\nonumber\\
\vec v&=&[-y,x,0],\\
\vec w&=&[xz,yz,-(x^{2}+y^{2})].\nonumber
\end{eqnarray}
The tetrad vectors are then given by \cite{brugmann08}
\begin{eqnarray}
&&n^{0}=\frac{1}{\sqrt{2}\alpha},\quad n^{i}=\frac{1}{\sqrt{2}}
(\frac{-\beta^{i}}{\alpha}-u^{i}),\nonumber\\
&&\ell^{0}=\frac{1}{\sqrt{2}\alpha},\quad\ell^{i}=\frac{1}{\sqrt{2}}
(\frac{-\beta^{i}}{\alpha}+u^{i}),\\
&&m^{0}=0,\qquad m^{i}=\frac{1}{\sqrt{2}}(v^{i}+iw^{i}).\nonumber
\end{eqnarray}
From the definition of $\Psi_4$ and Gauss-Codazzi-Mainardi
equations \cite{campanelli98,baker02}, we have
\begin{eqnarray}
\Psi_{4}&=&-{}^{(4)}R_{\alpha\beta\gamma\delta}n^\alpha\bar{m}^\beta n^\gamma
\bar{m}^\delta\nonumber\\
&=&-\frac{1}{2}[n^0n^0(R_{j\ell}-K_{jp}K^{p}_{\ell}+KK_{j\ell})
+4n^0n^kD_{[k}K_{\ell]j}\nonumber\\
&+&n^in^k(R_{ijk\ell}+2K_{i[k}K_{\ell]j})](v^j-iw^j)(v^\ell-iw^\ell).
\label{psi4def}
\end{eqnarray}
All the quantities in the last equation are physical ones in the
three-dimensional hypersurface. Let us relate them to the
conformal ones
\begin{eqnarray}
R_{ijk\ell}&=&e^{4\phi}(\tilde{R}_{ijk\ell}+4\tilde{\gamma}_{\ell[i}
\tilde{D}_{j]}\tilde{D}_{k}\phi-4\tilde{\gamma}_{k[i}\tilde{D}_{j]}
\tilde{D}_\ell\phi\nonumber\\
&-&16\phi_{,[i}\tilde{\gamma}_{j][k}\phi_{,\ell]}-8\tilde{\gamma}_{k[i}
\tilde{\gamma}_{j]\ell}\tilde{\gamma}^{mn}\phi_{,m}\phi_{,n}),\nonumber\\
R_{ij}&=&\tilde{R}_{ij}-2\tilde{D}_{i}\tilde{D}_{j}\phi-2\tilde{\gamma}_{ij}
\tilde{D}^{k}\tilde{D}_{k}\phi+4\phi_{,i}\phi_{,j}\nonumber\\
&-&4\tilde{\gamma}_{ij}\tilde{\gamma}^{mn}\phi_{,m}\phi_{,n},\nonumber\\
K_{ij}&=&e^{4\phi}(\tilde{A}_{ij}+\frac{1}{3}\tilde{\gamma}_{ij}K),\\
D_{i}K_{jk}&=&e^{4\phi}(\tilde{A}_{jk,i}-2\tilde{\Gamma}^\ell{}_{i(j}
\tilde{A}_{k)\ell}+\frac{1}{3}\tilde{\gamma}_{jk}K_{,i})\nonumber\\
&-&4K_{i(j}\phi_{,k)}+4\tilde{\gamma}_{i(j}K_{k)m}\tilde{\gamma}^{m\ell}
\phi_{,\ell}.\nonumber
\end{eqnarray}
Combining these equations together, we are able to compute
$\Psi_4$ on the entire three-hypersurface. When we transform
$m^\mu$ to $e^{i\zeta}m^\mu$, $\Psi_4$ will be transformed to
$e^{-2i\zeta}\Psi_4$ which results from Eq.~(\ref{psi4def}). The
functions with this kind of property are said to have spin weight
-2. Since the functions with different spin weight are orthogonal,
we can calculate the corresponding coefficients $A_{\ell m}$ with
respect to the spin weighted spherical harmonic function with spin
weight -2 \cite{wu07,brugmann08}, i.e., $Y^{-2}_{\ell m}$. The
major reason for choosing this type of function set is to reduce
numerical error during the calculations. The coefficients of the
gravitational radiation can be obtained from the integration of
$\Psi_4$ and each $Y^{-2}_{\ell m}$,
\begin{eqnarray}
A_{\ell m}=\langle Y^{-2}_{\ell m},\Psi_4\rangle\equiv\int^{2\pi}_{0}
\int^\pi_0\overline{Y^{-2}_{\ell m}}\Psi_4\sin\theta{\rm d}\theta{\rm d}\varphi,
\end{eqnarray}
where ${\ell}\geq 2$ and $-\ell\leq m\leq\ell$. The spin-weighted
spherical harmonics $Y^{-2}_{\ell m}$ can be defined in terms of
the Wigner d-functions (e.g., \cite{brugmann08}) as
\begin{eqnarray}
Y^{-2}_{\ell m}(\theta,\varphi)=\sqrt{\frac{2\ell+1}{4\pi}} {\rm
d}^\ell_{m2}(\theta)e^{im\varphi},
\end{eqnarray}
where
\begin{eqnarray}
{\rm
d}^\ell_{m2}(\theta)=\sum^{C_2}_{t=C_1}\frac{(-1)^t\sqrt{(\ell+m)!
(\ell-m)!}}{(\ell+m-t)!(\ell+2-t)!}\qquad\quad&&\nonumber\\
\frac{\sqrt{(\ell-2)!(\ell+2)!}}{(t-2-m)!t!}
(\cos\frac{\theta}{2})^{2\ell+m+2-2t}(\sin\frac{\theta}{2})^{2t-2-m},&&
\end{eqnarray}
with $C_1=\max(0,m-2)$ and $C_2=\min(\ell+m,\ell-2)$.
%%%%%%%%%%%%%%%%%%%%%%%%%%%%%%%%%%%%%
\section{The application of the unimodular determinant constraint}
%%%%%%%%%%%%%%%%%%%%%%%%%%%%%%%%%%%%%
\label{unimod} In the BSSN formulation the determinant of the
conformal metric is unimodular, i.e., Eq.~(\ref{detg1}). The first
spatial derivative of Eq.~(\ref{detg1}) gives
\begin{equation}\label{SEC}
\tilde{\gamma}^{ij}\tilde{\gamma}_{ij,k} = 0\quad\Longrightarrow\quad
\tilde{\Gamma}^j{}_{jk}= 0.
\end{equation}
We have imposed these three ``secondary'' constraints in the
calculation of the conformal Ricci curvature, $\tilde{R}_{ij}$ and
obtain its expression as in Eq.~(\ref{confricci}). The explicit
expression of the scalar curvature $\tilde R$ is
\begin{equation}\label{trR}
   \tilde{R}=-\frac{1}{2}\tilde{\gamma}^{ij}\tilde{\gamma}^{k\ell}
              \tilde{\gamma}_{ij,k\ell}+\tilde{\Gamma}^k{}_{,k}
  +\tilde{\Gamma}^{ijk}\left(2\tilde{\Gamma}_{jik}+\tilde{\Gamma}_{ijk}\right).
\end{equation}
Here the constraints (\ref{SEC}) have been used again.

The second spatial derivative of the unimodular determinant constraint gives
\begin{equation}
\tilde{\gamma}^{k\ell}\tilde{\gamma}_{k\ell,ij}
-4\tilde{\Gamma}_{k\ell i}\tilde{\Gamma}^{(k\ell)}{}_j = 0,
\end{equation}
and the trace of this ``tertiary'' constraint is
\begin{equation}\label{TER2}
\tilde{\gamma}^{ij}\tilde{\gamma}^{k\ell}\tilde{\gamma}_{ij,k\ell}
-4\tilde{\Gamma}_{ijk}\tilde{\Gamma}^{(ij)k} = 0.
\end{equation}
Using Eq.~(\ref{TER2}) we can rewrite Eq.~(\ref{trR}) to be
\begin{equation}\label{simtrR}
\tilde{R}= \partial_i\tilde{\Gamma}^i+\tilde{\Gamma}^{ijk}\tilde{\Gamma}_{jik}.
\end{equation}
Here the conformal trace curvature is only dependent on the
first derivatives of the conformal connection $\tilde{\Gamma}^i$
and of the conformal metric $\tilde{\gamma}_{ij}$.

The modified ADM mass volume integral in \cite{YHBS02}
\begin{eqnarray}
   M&=&\frac{1}{16\pi}\int_\Omega d^3x\bigg[e^{5\phi}
        \bigg(16\pi\rho + \tilde{A}_{ij}\tilde{A}^{ij}
        -\frac{2}{3}K^2\bigg)\nonumber\\
    &&\qquad\qquad\qquad-\tilde{\Gamma}^{ijk}\tilde{\Gamma}_{jik}
    + (1-e^\phi)\tilde{R}\bigg],\label{MADM}
\end{eqnarray}
can be understood better now with the expression (\ref{simtrR}).
Applying Eq.~(\ref{simtrR}) to Eq.~(\ref{MADM}) we have
\begin{eqnarray}
   M&=&\frac{1}{16\pi}\int_\Omega d^3x\bigg[e^{5\phi}
        \bigg(16\pi\rho + \tilde{A}_{ij}\tilde{A}^{ij}
        -\frac{2}{3}K^2\bigg)\nonumber\\
    &&\qquad\qquad\qquad+\partial_i\tilde{\Gamma}^i- e^\phi\tilde{R}\bigg].
\end{eqnarray}
For the cases in which $\tilde{\Gamma}^i$ is negligible, the ADM mass integral
expression returns to
\begin{equation}
   M = \frac{1}{16\pi}\displaystyle{\int_\Omega d^3x\left[e^{5\phi}
        \left(16\pi\rho + \tilde{A}_{ij}\tilde{A}^{ij} -\frac{2}{3}K^2\right)
        -e^\phi \tilde{R}\right]}
\end{equation}
%%%%%%%%%%%%%%%%%%%%%%%%%%%%%%%%%%%%%
\section{scalar field in a curved spacetime}\label{appendix_scalar}
%%%%%%%%%%%%%%%%%%%%%%%%%%%%%%%%%%%%%
\label{scalarguv}
The dynamics of a complex scalar field in a curved spacetime is described by
the following Lagrangian density \cite{liebling2007,das63}
\begin{equation}
\mathcal L=-\frac{1}{16\pi}R+\frac{1}{2}[\nabla_\mu\overline\Phi\nabla^\mu\Phi
+V(|\Phi|^2)],
\end{equation}
where $R$ is the scalar curvature, $\Phi$ is the scalar field, $\overline\Phi$
is its complex conjugate, and $V(|\Phi|^2)$ is the potential depending on
$|\Phi|^2$.
Here we consider the case where the potential takes the form
$V(|\Phi|^2)=m^2|\Phi|^2$ with $m$ being the mass parameter.
The action with the Lagrangian leads to Einstein's equation,
\begin{equation}
R_{\mu\nu}-\frac{1}{2}g_{\mu\nu}R=8\pi T_{\mu\nu},
\end{equation}
where $R_{\mu \nu}$ is the Ricci tensor and $T_{\mu \nu}$ is the stress-energy
tensor defined by
\begin{equation}
T_{\mu\nu}=\nabla_{(\mu}\Phi\nabla_{\nu)}\overline\Phi-\frac{1}{2}g_{\mu\nu}
[\nabla_\sigma\Phi\nabla^\sigma\overline\Phi+m^2|\Phi|^2].
\end{equation}
The matter source terms in the BSSN formulation are constructed
from the stress-energy tensor via the expressions
\cite{modified_3+1}
\begin{eqnarray}
\rho  &=&n_\mu n_\nu T^{\mu\nu}, \\
s_{ij}&=&\gamma_{i\mu}\gamma_{j\nu}T^{\mu\nu},\\
s_i   &=&-\gamma_{i\mu}n_\nu T^{\mu\nu},\\
s     &=&\gamma^{ij}s_{ij}.
\end{eqnarray}
The field equation for a {\it massless} scalar field is
\begin{equation}
\nabla_\mu\nabla^\mu\Phi = 0,
\end{equation}
which can be expanded as
\begin{equation}
\nabla_\mu\nabla^\mu\Phi=\frac{1}{\sqrt{-g}}\partial_\mu(\sqrt{-g}g^{\mu\nu}
\partial_\nu\Phi)=0.
\end{equation}
With the metric given by
\begin{equation}
ds^2=-\alpha^2dt^2+\gamma_{ij}(dx^i+\beta^idt)(dx^j+\beta^jdt),
\end{equation}
we can re-write the field equation and decompose this second-order equation
into two first-order-in-time equations:
\begin{eqnarray}
(\partial_t-\beta^i\partial_i)\Phi&=&\alpha\Pi,\label{scalar1}\\
(\partial_t-\beta^i\partial_i)\Pi&=&e^{-4\phi}\alpha[{\tilde\gamma}^{ij}
\Phi_{,ij}-({\tilde\Gamma}^i-2
{\tilde\gamma}^{ij}\phi_{,i})\Phi_{,j}]\nonumber\\
&+&e^{-4\phi}{\tilde\gamma}^{ij}\alpha_{,i}\Phi_{,j}+\alpha K\Pi.
\label{scalar2}
\end{eqnarray}
Combining Eqs.~(\ref{scalar1}), (\ref{scalar2}) with Einstein's
equation, we can study the dynamical interaction between a
massless scalar field and the curved space-time where the scalar
field lives via the numerical analysis.
%%%%%%%%%%%%%%%%%%%%%%%%%%%%%%%%%%%%%%%%%%

\end{document}